\documentclass[pra,aps,twocolumn,showpacs,groupedaddress,superscriptaddress]{revtex4-1}
\usepackage{amsmath,amsfonts,amssymb,graphics,graphicx,epsfig,color,times}

\usepackage[utf8x]{inputenc}
\usepackage{color}
\usepackage{bbm} 

\usepackage{amsfonts,amsmath,amssymb,stmaryrd}

\usepackage{graphicx}

\usepackage{subfigure}  

\usepackage{bbm} 

\usepackage{hyperref}

\usepackage{epsfig}

\usepackage{mathrsfs}

\usepackage{verbatim}

\usepackage{xpatch}

\makeatletter
\patchcmd{\@ssect@ltx}
    {\addcontentsline{toc}{#1}{\protect\numberline{}#8}}
    {}
    {}
    {}
\makeatother

\renewcommand{\l}{\left(}

\renewcommand{\r}{\right)}

\newcommand{\bra}[1]{\langle#1|}

\newcommand{\ket}[1]{|#1\rangle}

\renewcommand{\H}{\hat{\mathcal{H}}}
\renewcommand{\c}{\hat{c}}
\renewcommand{\a}{\hat{a}}

\newcommand{\cd}{\hat{c}^\dagger}
\newcommand{\ad}{\hat{a}^\dagger}
\newcommand{\bd}{\hat{b}^\dagger}
\renewcommand{\b}{\hat{b}}
\newcommand{\dd}{\hat{d}^\dagger}
\renewcommand{\d}{\hat{d}}

\newcommand{\g}{\hat{\gamma}}

\newcommand{\hc}{\text{h.c.}}

\usepackage{ulem}	

\newcommand{\nocontentsline}[3]{}
\newcommand{\tocless}[2]{\bgroup\let\addcontentsline=\nocontentsline#1{#2}\egroup}

\usepackage{array}

\usepackage{bm}	
\renewcommand{\vec}[1]{\bm{#1}}

\newcommand{\Zak}{\text{Zak}}
\newcommand{\TP}{\text{TP}}

\begin{document}
\normalem	

\title{Topological polarons, quasiparticle invariants and their detection in 1D symmetry-protected phases}

\author{F. Grusdt}
\email[Corresponding author email: ]{fabian.grusdt@tum.de}
\affiliation{Department of Physics and Institute for Advanced Study, Technical University of Munich, 85748 Garching, Germany}
\affiliation{Munich Center for Quantum Science and Technology (MCQST), Schellingstr. 4, D-80799 M\"unchen, Germany}

\author{N. Y. Yao}
\affiliation{Department of Physics, University of California Berkeley, Berkeley, CA 94720, USA}

\author{E. A. Demler}
\affiliation{Department of Physics, Harvard University, Cambridge, Massachusetts 02138, USA}

\date{\today}

\begin{abstract}
In the presence of symmetries, one-dimensional quantum systems can exhibit topological order, which in many cases can be characterized by a quantized value of the many-body geometric Zak or Berry phase. We establish that this topological Zak phase is directly related to the Zak phase of an elementary quasiparticle excitation in the system. By considering various  systems, we establish this connection for a number of different interacting phases including: the Su-Schrieffer-Heeger model, p-wave topological superconductors, and the Haldane chain. Crucially, in contrast to the bulk many-body Zak phase associated with the ground state of such systems, the topological invariant associated with quasiparticle excitations (above this ground-state) exhibit a more natural route for direct experimental detection. To this end, we build upon recent work [Nature Communications 7, 11994 (2016)] and demonstrate that mobile quantum impurities can be used, in combination with  Ramsey interferometry and Bloch oscillations, to directly measure these quasiparticle topological invariants. Finally, a concrete experimental realization of our protocol for dimerized Mott insulators in ultracold atomic systems is discussed and analyzed.
\end{abstract}

\maketitle

\tableofcontents

\section{Introduction}
\begin{figure}[t!]
\centering
\epsfig{file=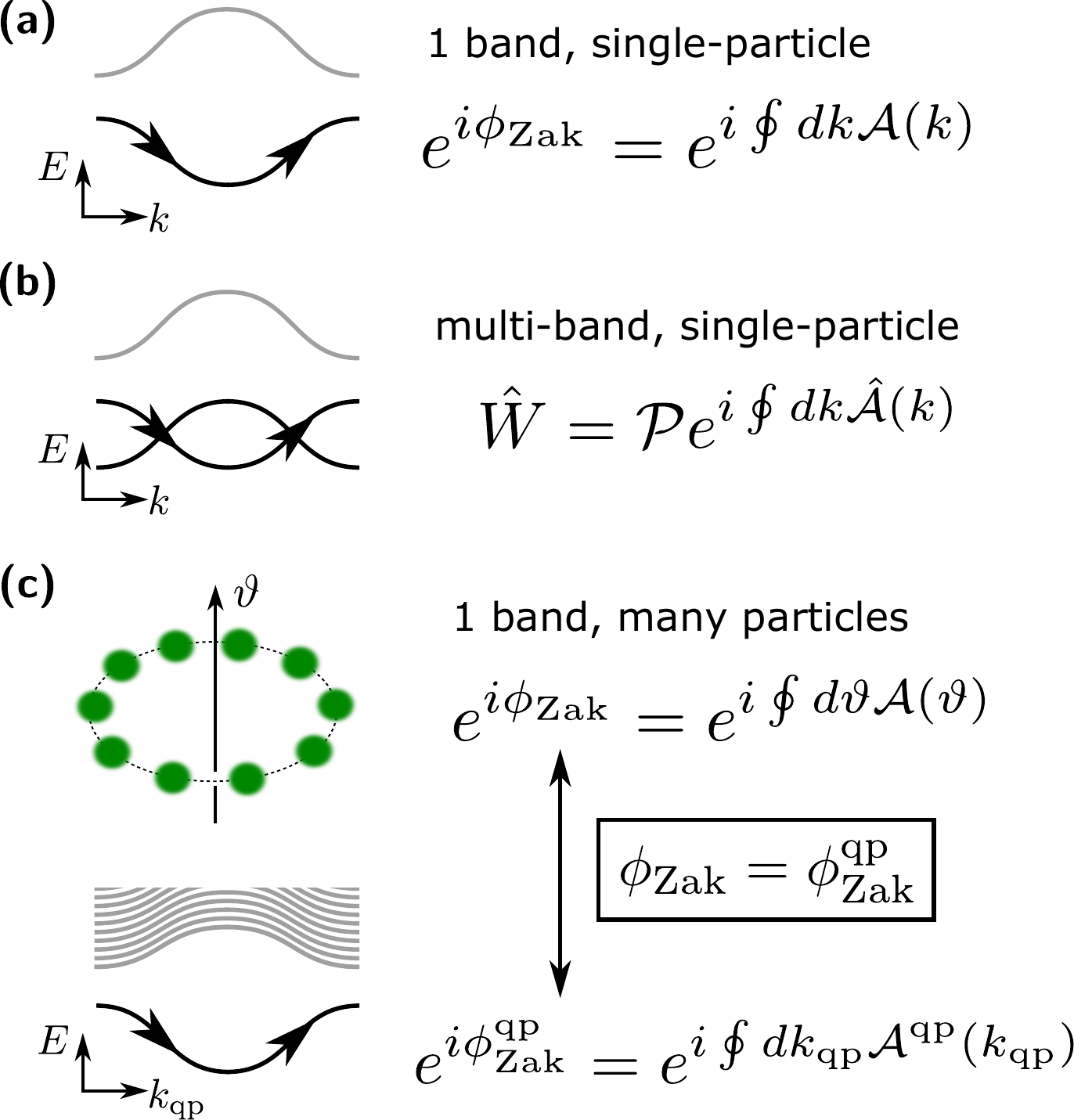, width=0.43\textwidth}
\caption{Using a combination of Ramsey interferometry and Bloch oscillations, the Berry or Zak phase of a Bloch band can be directly measured. This approach was demonstrated for one-band \cite{Atala2012} (a) and multi-band systems \cite{Li2016} (b) where single-particle effects have been observed. Here we generalize the approach to strongly interacting one-dimensional systems, where topological order can be characterized by the many-body Zak phase (c). To measure it, we make use of its relationship to the Zak phase of a quasiparticle excitation in the many-body system.}
\label{fig:Intro}
\end{figure}

Developments in the quantum control of individual atoms, ions, molecules and photons have led to the exciting ability to realize certain topological phases of matter in ultracold quantum simulators \cite{Lin2009,Gemelke2010,Struck2012,Atala2012,Aidelsburger2013,Miyake2013,Rechtsman2013a,Verbin2013,Hafezi2013a,Jotzu2014,Aidelsburger2014,Ningyuan2015,Stuhl2015,Mancini2015,Schine2016,Tai2017,Lohse2018,Leseleuc2018arxiv}. 
One of the crucial new features of such systems is the ability to directly measure the non-local topological invariants that underly these phases of matter, enabling in principle, the direct experimental classification of topological phases.
In this context, a particularly powerful approach has emerged, which combines Bloch oscillations and Ramsey interferometry in order to measure topological invariants in ultracold atomic systems \cite{Atala2012}.
The essence of this approach is summarized in in Fig.~\ref{fig:Intro} (a): A particle can be moved through the Brillouin zone to directly measure the geometric Berry \cite{Berry1984} or Zak \cite{Zak1989} phase characterizing the underlying bandstructure. This approach has been generalized to  multi-band systems \cite{Grusdt2014Z2,Li2016,Flaschner2016}, where the Wilson loop can be directly measured using similar techniques [Fig.~\ref{fig:Intro} (b)], as well as to two-dimensional systems \cite{Abanin2012,Duca2014} and quantum random walks \cite{Kitagawa2010a,Ramasesh2017,Flurin2017}. 
More recently, a tremendous amount of attention has focused on extending these interferometric schemes, as well as alternative approaches \cite{Tran2017,Ma2017,Tarnowski2018,Asteria2018}, beyond single particle bandstructures to the measurement of many-body topological invariants \cite{Grusdt2016TP,Repellin2018,Zheng2018}.

In this article, we present three main results related to this broader goal. First, building upon the protocol introduced in Ref.~\cite{Grusdt2016TP}, we demonstrate that interferometry can be used to measure the many-body Zak phase of one-dimensional symmetry-protected topological (SPT) matter \cite{Ryu2010,Kitaev2009,Chen2011,Chen2011b,Turner2011,Fidkowski2011,Pollmann2012}. Second, we demonstrate that, in certain models, one can define a topological invariant associated with the quasiparticle excitations above the ground state, and that these excitations effectively inherit the ground state's topology.
Finally, we propose an experimental realization of our protocol that can be implemented in near-term ultracold atomic systems. 
Before jumping into the details, we provide an intuitive blueprint for how to understand our results.

The starting point of our work is always a gapped phase with symmetry-protected topological order, in an interacting one-dimensional quantum many-body system (i.e.~a 1D SPT phase). While there exists a plethora of such phases \cite{Chen2011b,Turner2011,Fidkowski2011}, we will focus on cases that can be characterized by a quantized many-body Zak  phase of the ground state manifold. The latter is defined by introducing twisted boundary conditions \cite{Niu1985} in a periodic system, which makes a direct experimental measurement extremely challenging and requires intensive ground state degeneracy.

Instead of facing this challenge, we consider a single quasiparticle excitation above the ground state, which carries a well-defined quantum number, e.g. spin or charge. 
The set of all many-body states with exactly one such excitation form a band at low energies, which can be labeled by the momentum of the quasiparticle. 
We will further assume that this quasiparticle band is separated by a gap from all other bulk excitations. 
Interestingly, it has recently been argued that this scenario is generic for strongly interacting systems in one spatial dimension \cite{Verresen2018}.

To define a topological invariant characterizing the 1D SPT phase, we propose to treat the low energy quasiparticle band analogous to the conventional bandstructure of a single-particle excitation. 
This allows one to naturally define the Zak phase of the quasiparticle; however, the crucial difference is that the underlying quantum mechanical wavefunction is defined on the high-dimensional many-body Hilbert space.

A priori, it is not obvious how the Zak phase of the quasiparticle excitation, $\phi_{\rm Zak}^{\rm qp}$, relates to the many-body Zak phase of the bulk ground state without the quasiparticle. We will show below that in many cases of interest the many-body Zak phase, defined by twisted periodic boundary conditions, gives the same result as the newly defined quasiparticle Zak phase. 
The connection is provided by a theorem by King-Smith and Vanderbilt \cite{Kingsmith1993}, who showed that the Zak phase $\phi_{\rm Zak}$ of a single particle is directly related to the polarization $P \propto \phi_{\rm Zak}$. In a single-particle band, the polarization is determined by the center-of-mass of its Wannier functions $P = \bra{w} \hat{x} \ket{w}$, whereas the many-body polarization \cite{Ortiz1994} is defined by the center-of-mass of the many-body system, $P = \langle \hat{X} \rangle$. Here we use the notion of polarization in a general sense, as it can be related to any quantum number, such as spin or charge. 

Since we interpret the quasiparticle band as a single-particle excitation, its Zak phase $\phi_{\rm Zak}^{\rm qp}$ describes the center-of-mass of its effective Wannier function, defined in the many-body Hilbert space. This Wannier function reflects the spatial structure of the correlated many-body state locally, since the system is gapped and hence has a finite correlation length. Since this local structure in the bulk also directly relates to the many-body Zak phase $\phi_{\rm Zak}$ of the state, we expect that the latter is generically related to the quasiparticle Zak phase $\phi_{\rm Zak}^{\rm qp}$. To make these arguments precise, we consider specific models and establish case by case that the quasiparticle and many-body Zak phases are equivalent, $\phi_{\rm Zak}^{\rm qp} = \phi_{\rm Zak}$.

To measure the topological invariant $\phi_{\rm Zak}^{\rm qp}$ of the quasiparticle band, we follow the approach from Ref.~\cite{Grusdt2016TP} and introduce a mobile quantum impurity. We assume that its interactions with the many-body system lead to the formation of a bound state between the impurity and the quasiparticle. In this bound state, which we refer to as a topological polaron \cite{Grusdt2016TP}, the mobile impurity inherits the topological properties of the surrounding many-body system. In the strong coupling regime, we show that the resulting Zak phase of the topological polaron $\phi_{\rm Zak}^{\rm TP} = \phi_{\rm Zak}^{\rm qp}$ is equivalent to the quasiparticle invariant. Using an impurity with two internal (pseudo-) spin states allows to address the impurity and measure the Zak phase by the same techniques developed for non-interacting particles \cite{Atala2012}, which we review later in this article. 

The topological polarons discussed in this article constitute an example how interactions of a mobile quantum impurity with a topologically non-trivial many-body system can lead to the formation of a new quasiparticle which inherits the topological properties of the surrounding bath. Here we focus on the case where the impurity binds to an additional quasiparticle excitation \cite{Grusdt2016TP}. Recently a similar situation has been discussed where a mobile impurity is dressed with particle-hole excitations of a Chern insulator \cite{CamachoGuardian2019}, resembling the formation of fermi polarons \cite{Schirotzek2009,Koschorreck2012,kohstall2012metastability} but in the presence of non-trivial band topology.

This article is organized as follows. We will begin by giving a brief overview of the method in Sec.\ref{sec:Overview}. In Sec.\ref{sec:BIs} we introduce some necessary theoretical background and define the quasiparticle Zak phase for non-interacting fermionic systems. In Sec.\ref{sec:MIs} we return to interacting many-body systems and discuss dimerized Mott insulators of bosons in one dimension. A concrete experimental setup is suggested, for which the required protocol is discussed in detail and exact numerical results are presented. In Sec.\ref{sec:OtherSystems} we discuss how the method can be applied to detect topological order in other physical systems, including frustrated spin chains and topological superconductors. We close with a summary and outlook in Sec.\ref{sec:SummaryOutlook}.

\section{Overview}
\label{sec:Overview}
In many cases, topological invariants of correlated systems can be formulated in terms of twisted periodic boundary conditions \cite{Niu1985}, see Fig.~\ref{fig:Intro} (c). By adiabatically changing the phase $\vartheta$, picked up by the system when a particle is taken around the system once, the many-body wavefunction acquires a geometric phase, the many-body Zak phase $\phi_{\rm Zak}$. The effect of changing $\vartheta$ in the many-body system is similar to the effect of a force acting on a single particle in a lattice, as a consequence of which the particle picks up the Zak phase of the occupied Bloch band \cite{Zak1989,Atala2012}.  

Because detecting the overall phase of a many-body wavefunction, and realizing twisted periodic boundary conditions, are practically impossible experimentally, a direct measurement of the many-body Zak phase is extremely challenging. The key idea of the scheme for the measurement of many-body topological invariants \cite{Grusdt2016TP} is to utilize the relationship between the many-body Zak phase $\phi_{\rm Zak}$ and the Zak phase $\phi_{\rm Zak}^{\rm qp}$ characterizing the effective bandstructure of a quasiparticle excitation with momentum $k_{\rm qp}$, 
\begin{equation}
\phi_{\rm Zak}^{\rm qp} = \int_{\rm qp-BZ} dk_{\rm qp}~ \underbrace{\bra{\psi(k_{\rm qp})} i \partial_{k_{\rm qp}} \ket{\psi(k_{\rm qp})}}_{=\mathcal{A}^{\rm qp}(k_{\rm qp})},
\end{equation}
discussed above, see also Fig.~\ref{fig:Intro} (c). 

In this article we will establish a one-to-one relation between the two, $\phi_{\rm Zak}^{\rm qp} = \phi_{\rm Zak}$, for a various models exhibiting symmetry-protected topological order. The direct measurement of $\phi_{\rm Zak}^{\rm qp}$ can then be achieved by binding a mobile impurity, acting as a coherent probe, to the quasiparticle and applying the interferometric schemes \cite{Atala2012,Abanin2012,Grusdt2014Z2,Duca2014} developed for non-interacting particles, which we review below. 

\begin{figure}[t]
\centering
\epsfig{file=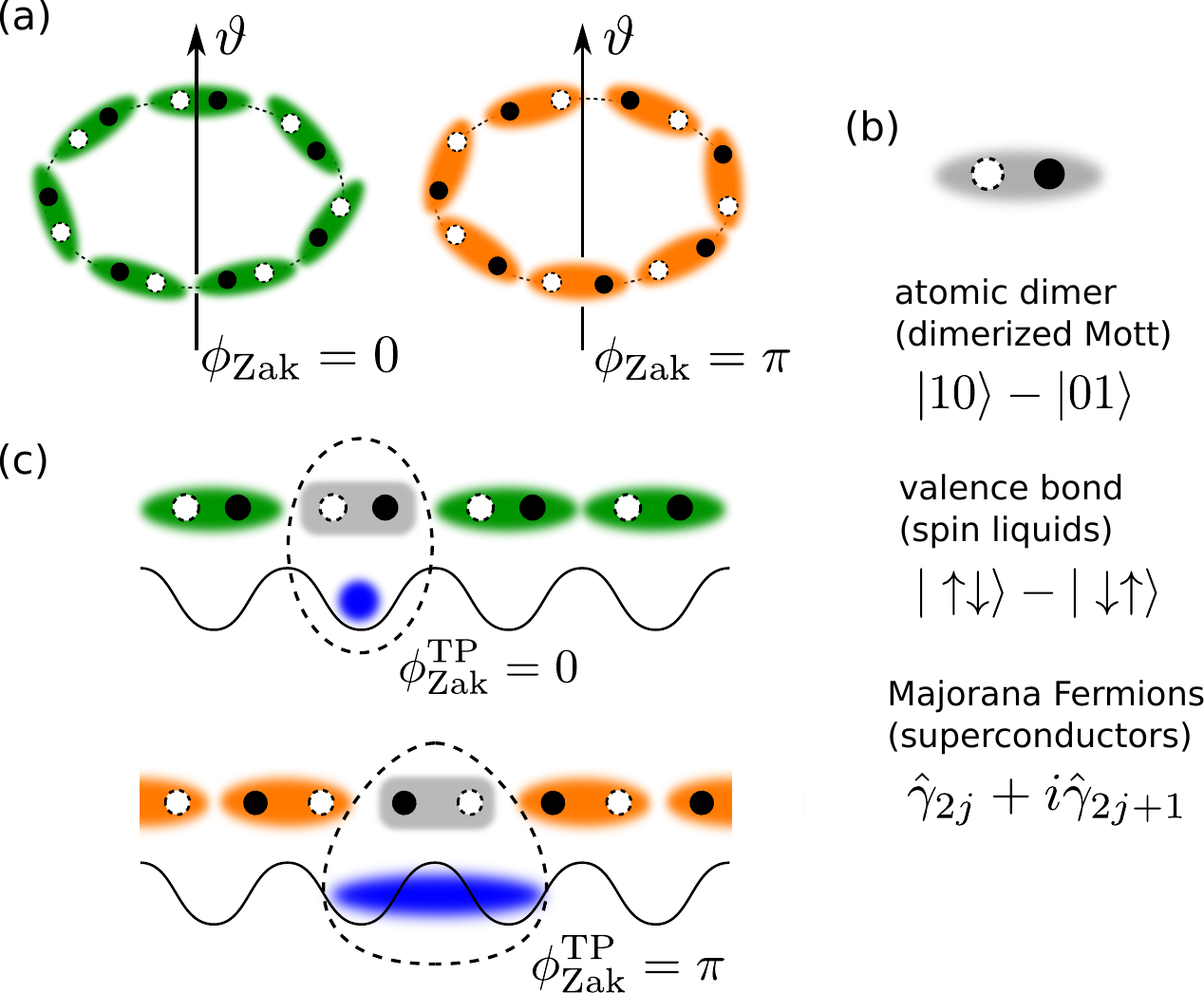, width=0.47\textwidth}
\caption{Many symmetry-protected topologically ordered phases in one dimension can be distinguished by their dimerization patterns. (a) Different dimer configurations give rise to different values of the many-body Zak phase $\phi_\Zak$. It is obtained by varying the twisted periodic boundary conditions, which can also be understood as an adiabatic change of a Aharonov-Bohm phase $\vartheta$ picked up when particles encircle the periodic system once. Dimers can be realized in various systems (b), where different symmetries give rise to a quantization of the many-body Zak phase to $\phi_\Zak=0,\pi$. (c) To measure the many-body Zak phase we couple a mobile impurity (blue) to an elementary topological excitation of the many-body system, corresponding e.g. to a broken dimer. In this way a topological polaron (TP) is formed. When the impurity lattice has the same period as the dimer covering, the Zak phase of the TP $\phi_\Zak^{\rm TP}$ allows to distinguish topologically inequivalent states.}
\label{fig:overview}
\end{figure}

The scheme can be applied to a wide range of systems with symmetry-protected topological order in one dimension. Examples discussed in this article include dimerized Mott insulators \cite{Muth2008,Nascimbene2012,Grusdt2013EdgeStates,Chen2013a,Jurgensen2014}, anti-ferromagnetic spin chains \cite{Majumdar1969a,Haldane1983b,Affleck1987} and topological superconductors \cite{Read1999,Ryu2002,Kitaev2001}, see Fig.\ref{fig:overview} (a), (b). This moreover paves the way for measurements of topologically invariant Berry phases characterizing gapped quantum spin liquids \cite{Hatsugai2006,Hatsugai2007}, possibly also in higher dimensions.

To measure the many-body Zak phase directly we suggest to couple a two-component mobile impurity (pseudospins $\uparrow$, $\downarrow$) to a topological excitation of the many-body system as in Ref.\cite{Grusdt2016TP}, see Fig.\ref{fig:overview} (c). Then a similar interferometric scheme as implemented in Ref.~\cite{Atala2012} can be applied to map out the topology of the resulting impurity wavefunction. The key idea in Ref.~\cite{Atala2012} was to combine Ramsey interferometry with Bloch oscillations to measure the Zak phase
\begin{equation}
\varphi_{\rm Zak}^\alpha = \int_{\rm BZ} dk~ \underbrace{\bra{u_\alpha(k)} i \partial_k \ket{u_\alpha(k)}}_{=\mathcal{A}_\alpha(k)}
\label{eq:phiZakUk}
\end{equation}
of a Bloch band $\ket{u_\alpha(k)}$. When opposite forces are applied to the two pseudospin components of an impurity in $\ket{u_\alpha(k)}$, they undergo Bloch oscillations in opposite directions. The relative phase picked up by the different spin components after crossing half the Brillouin zone (BZ) is equal to the Zak phase of the Bloch band.

The impurity serves as a coherent probe of the host many-body system. When it is coupled to elementary (topological) excitation a quasiparticle is formed which we call a topological polaron (TP) \cite{Grusdt2016TP}, see also Fig.\ref{fig:overview}. From the impurity, the TP inherits two pseudospin components which are used for Ramsey interferometry. From the elementary excitation in the many-body system, on the other hand, the TP inherits its topological properties which we want to detect. Now the key idea of the protocol is to measure the Zak phase characterizing the band structure of a single TP $\ket{\psi_\TP(k)}$,
\begin{equation}
\phi_{\rm Zak}^{\TP} =  \int_{\TP - \rm BZ} dk~ \bra{\psi_\TP(k)} i \partial_k \ket{\psi_\TP(k)},
\label{eq:defTPzak}
\end{equation}
in analogy to Eq.\eqref{eq:phiZakUk}. We consider situations where the impurity is strongly coupled to the quasiparticle, which ensures that the topology of the TP is dictated by the the quasiparticle topology, i.e. $\phi_{\rm Zak}^{\TP} = \phi_{\rm Zak}^{\rm qp}$, see Ref.~\cite{Grusdt2016TP} for a general discussion. Although for the theoretical analysis we focus on the case of a single impurity, the results carry over to situations of sufficiently low impurity concentration (their mutual interactions should be negligible).

We propose a concrete experimental realization of the scheme with ultracold atoms. To this end we consider the half-filling Mott insulating (MI) phase of the Bose-Hubbard model for a Rice-Mele lattice \cite{Rice1982}. This system has recently been realized with ultracold atoms \cite{Lohse2015}. For regions in parameter space where the Hamiltonian is invariant under spatial inversion, it has symmetry-protected topological order characterized by a quantized many-body Zak phase and topological edge states \cite{Grusdt2013EdgeStates}. In addition the model realizes a topological Thouless pump \cite{Thouless1983,Berg2011,Hayward2018a} where a quantized amount of charge is pumped in each cycle \cite{Lohse2015,Nakajima2016}. It is related to the many-body Chern number $\mathcal{C}$, defined as a winding of the many-body Zak phase along the loop in parameter space. We demonstrate that both the quantized many-body Zak phases as well as the Chern number characterizing the Thouless pump can be directly measured using TPs.

\section{Theoretical background -- band insulators}
\label{sec:BIs}
\begin{figure}[b!]
\centering
\epsfig{file=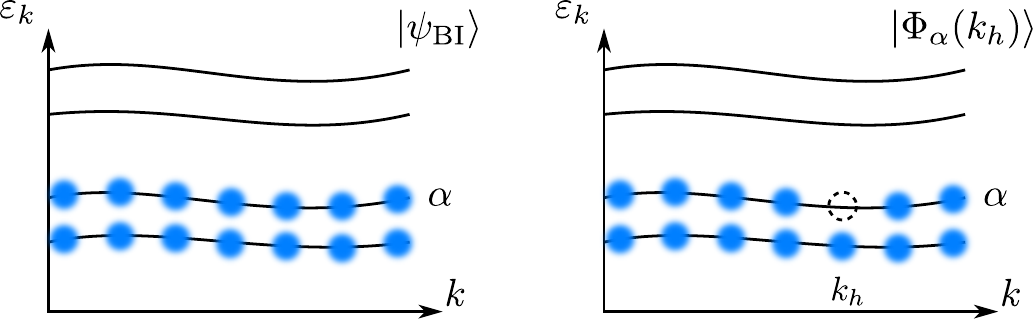, width=0.5\textwidth}
\caption{We consider a band insulator of non-interacting fermions (left). Its many-body Zak phase is equivalent to the sum of all Zak phases of hole excitations $\ket{\Phi_\alpha}$ shown on the right.}
\label{fig:BIs}
\end{figure}

We begin by discussing many-body Zak phases $\phi_\Zak$ for band insulators of free fermions in one dimension. When $\cd_{\alpha,k}$ creates a fermion in band $\alpha$ and at quasimomentum $k$ in the Brillouin zone, the band insulator state can be written as
\begin{equation}
\ket{\psi_{\rm BI}} = \prod_{\alpha ~\rm occ.} ~ \prod_{k \in \rm BZ} ~ \cd_{\alpha,k} \ket{0},
\label{eq:BIdef}
\end{equation}
see Fig.\ref{fig:BIs}. Here the product $\prod_\alpha$ includes all occupied bands and we consider periodic boundary conditions. 

Now we calculate the many-body Zak phase for the band insulator and review how it is related to the Zak phase, Eq.\eqref{eq:phiZakUk}, of the underlying Bloch wavefunctions $\ket{u_\alpha(k)}$. Then we generalize the calculation for single hole excitations of the band insulator.

\subsection{Many-body Zak phase and twisted boundary conditions}
\label{subsec:MBzakAndTwistedBCs}
The many-body Zak phase characterizing the band insulator $\ket{\psi_{\rm BI}}$ can be defined by introducing twisted periodic boundary conditions \cite{Niu1985,Ortiz1994}, see Fig.\ref{fig:overview} (a). For all particles $j=1,...,N$, with $N$ denoting the total particle number and $L$ the system size, it holds
\begin{equation}
\psi_{\rm BI}(x_1,...,x_j+L,...,x_N) = e^{i \vartheta} \psi_{\rm BI}(x_1,...,x_j,...,x_N).
\label{eq:twistedBCs}
\end{equation}
When the twist angle $\vartheta$ is varied adiabatically from $0$ to $2 \pi$, the gapped band insulator state $\ket{\psi_{\rm BI}(\vartheta)}$ returns to itself because a $2 \pi$ phase corresponds to a pure gauge transformation $\hat{U}$ \cite{Laughlin1981}. The wavefunctions $\ket{\psi_{\rm BI}(0)}$ before and $\ket{\psi_{\rm BI}(2 \pi)}$ after introducing $2 \pi$ twist are thus related by a global gauge transformation, 
\begin{equation}
\ket{\psi_{\rm BI}(2 \pi)} = e^{i \phi_{\rm Zak}^{\rm BI}} \hat{U} \ket{\psi_{\rm BI}(0)},
\end{equation}
up to a phase $\phi_{\rm Zak}^{\rm BI}$ which defines the many-body Zak phase \cite{Zak1989,Ortiz1994}. When a gauge choice is made where $\hat{U}=1$, it simply reads
\begin{equation}
\phi_{\rm Zak}^{\rm BI} =  \int_0^{2 \pi} d\vartheta~ \bra{\psi_{\rm BI}(\vartheta)} i \partial_\vartheta \ket{\psi_{\rm BI}(\vartheta)}.
\label{eq:defPhiZak}
\end{equation}

Inversion symmetry has been shown to lead to a quantization of the many-body Zak phase to values $\phi_\Zak = 0, \pi$ \cite{Zak1989,Atala2012,Grusdt2013EdgeStates}. This is one example for symmetry-protected topological order characterized by the Zak phase, but other symmetries can also be sufficient for a quantization of $\phi_\Zak$, see e.g. Ref. \cite{Hatsugai2006,Ryu2010}.

To gain better understanding of the effect of twisted boundary conditions on the band insulator \eqref{eq:BIdef}, we consider the single-particle Bloch states for $\vartheta=0$,
\begin{equation}
\Psi_{\alpha,k}(x) = e^{i k x} u_{\alpha,k} (x), \qquad k = \frac{2 \pi}{L}n, \qquad n \in \mathbb{Z},
\end{equation}
which fulfill periodic boundary conditions, i.e. $\Psi_{\alpha,k}(x+L)  = \Psi_{\alpha,k}(x)$. As usual we make the gauge choice $\Psi_{\alpha,k}(x)=\Psi_{\alpha,k+2 \pi/L}(x)$. To construct the corresponding eigenfunctions for twisted boundary conditions $\vartheta\neq0$ we displace the quasimomentum by $\vartheta/L$,
\begin{equation}
\Psi_{\alpha,k}(\vartheta , x) = e^{i \l k + \vartheta / L \r x} ~ u_{\alpha,k + \vartheta / L  } (x).
\label{eq:defPsikTheta}
\end{equation}
Thus a time-dependent twist angle corresponds to a force $F=-\dot{\vartheta}/L$ acting on the fermions in the BI. The wavefunctions \eqref{eq:defPsikTheta} have the property that $\Psi_{\alpha,k}(\vartheta , x+L) = e^{i \vartheta} \Psi_{\alpha,k}(\vartheta , x)$, and one easily checks that they are proper eigenfunctions of the lattice Hamiltonian. Moreover they satisfy the gauge convention $\Psi_{\alpha,k+2 \pi / L}(\vartheta , x) = \Psi_{\alpha,k}(\vartheta , x)$. Most importantly, we observe that a $2 \pi$ twist of $\vartheta$ results in an adiabatic change of the momentum by $\Delta k = 2 \pi/L$.

Now we use the Bloch wavefunctions \eqref{eq:defPsikTheta} to calculate the many-body Zak phase $\phi_\Zak^{\rm BI}$ of the band insulator \eqref{eq:BIdef}, i.e. the Slater determinant state constructed from $\Psi_{\alpha,k}(\vartheta,x)$. We obtain
\begin{equation}
\ket{\psi_{\rm BI}(\vartheta)} = e^{i \phi_{\Zak}^{\rm BI}} e^{i \vartheta \sum_{j=1}^N x_j / L} \ket{\psi_{\rm BI}(0)};
\label{eq:gaugeChoiceBI}
\end{equation}
When $\vartheta$ is a multiple of $2 \pi$, the exponential $e^{i \vartheta \sum_{j=1}^N x_j / L}=\hat{U}(\vartheta)$ is a pure gauge transformation. The many-body Zak phase can now be calculated from
\begin{multline}
\phi_\Zak^{\rm BI} = \int_0^{2 \pi} d \vartheta  ~ \bra{\psi_{\rm BI}(\vartheta)} \hat{U}(\vartheta) i \partial_\vartheta \hat{U}^\dagger(\vartheta) \ket{\psi_{\rm BI}(\vartheta)} \\
=  \sum_{\alpha ~ \rm occ.} \sum_{k \in \rm BZ} \int_k^{k+2 \pi/L} dq~ \bra{u_\alpha(q)} i \partial_q \ket{u_\alpha(q)},
\label{eq:defZ}
\end{multline}
i.e. every fermion moves adiabatically from $k$ to $k + 2 \pi/L$ as $\vartheta$ is varied from $0$ to $2 \pi$. Together all fermions from one band $\alpha$ pick up its Zak phase, cf. Eq.\eqref{eq:phiZakUk}, and we arrive at the expression
\begin{equation}
\phi_\Zak^{\rm BI} =  \sum_{\alpha ~ \rm occ.} \varphi_\Zak^\alpha.
\end{equation}

\subsection{Zak phases of hole excitations}
Next we consider hole excitations in the band insulator, and show that their Zak phases are directly related to the many-body Zak phase of the hosting band insulator. This one-to-one relation is at the heart of the interferometric measurement scheme, where the Zak phase of the hole is detected by coupling it to a mobile impurity which serves as a coherent probe.

Hole excitations in the band insulator exist in every band $\alpha$ and we can write them as (see also Fig.\ref{fig:BIs})
\begin{equation}
\ket{\Phi_\alpha(k)} = \c_{\alpha,k} \ket{\psi_{\rm BI}}.
\end{equation}
To define the Zak phase of the hole as in Zak's original paper \cite{Zak1989}, a force $F$ should be applied directly to the hole. Then its quasimomentum changes adiabatically in time, $k(t)=k-F t$. After completing a full Bloch cycle in the Brillouin zone the hole wavefunction $\ket{\Phi_\alpha(k)}$ returns to itself up to a gauge transformation and a phase factor $e^{i \phi_\Zak^{h}(\alpha)}$ defining the Zak phase of the hole excitation $\phi_\Zak^{h}(\alpha)$.

To make use of the twisted periodic boundary conditions discussed in the previous section, we include the force $F$ acting on the hole. Projecting it onto the occupied bands
we can write it as
\begin{equation}
\H_F = - \sum_{\alpha ~ \rm occ.} \sum_j F X_{j,\alpha} \l  1 - \hat{n}_j^\alpha \r.
\label{eq:defFhole}
\end{equation}
Here $X_{j,\alpha} = \bra{w_j^\alpha} \hat{x} \ket{w_j^\alpha}$ denotes the center of mass of the Wannier function $\ket{w_j^\alpha}$ at site $j$ corresponding to band $\alpha$ ($\hat{x}$ is the position operator).

The second term in Eq.\eqref{eq:defFhole} describes a force $-F$ acting on all fermions in the BI. To understand its effect on the hole state $\ket{\Phi_\alpha(k)}$ we apply the force for one Bloch period $T=2 \pi / a F$. Formulated in terms of twisted boundary conditions on a torus this corresponds to $L/a$ full twists, $\Delta \vartheta = -2 \pi L/a$. From the calculations in the proceeding section we know that during this process every fermion picks up a contribution $-\varphi_\Zak^\beta$ to the many-body Zak phase. Because the hole corresponds to a missing fermion in the BI, the geometric phase due to the second term in Eq.\eqref{eq:defFhole} reads 
\begin{equation}
\phi_2= - \frac{L}{a} \sum_{\beta ~ \rm occ.} \varphi_\Zak^\beta + \varphi_\Zak^\alpha.
\label{eq:phi2Res}
\end{equation} 
Here we assumed for simplicity that hole bands are all separated by gaps, but the argument can be generalized.

The first term in Eq.\eqref{eq:defFhole} adds an additional geometric phase given by $\phi_1 = T F \sum_{j} \sum_{\beta ~ \rm occ.} X_{j,\beta}$. To calculate this term, we make use of a theorem by King-Smith and Vanderbilt \cite{Kingsmith1993} which relates the Wannier center $X_{j,\beta}$ to the Zak phase,
\begin{equation}
X_{j,\beta} = \bra{w_j^\beta} \hat{x} \ket{w_j^\beta} = \frac{a}{2 \pi} \varphi_\Zak^\beta.
\end{equation}
Using this expression we obtain
\begin{equation}
\phi_1 = \sum_j \sum_{\beta~ \rm occ.} \varphi_\Zak^\beta = \frac{L}{a} \sum_{\beta~ \rm occ.} \varphi_\Zak^\beta.
\end{equation}
This term exactly cancels the first term in Eq.\eqref{eq:phi2Res}.

Combining our results we conclude that the Zak phase of the hole $\phi_\Zak^{h}(\alpha) = \phi_1 + \phi_2 $ is given by
\begin{equation}
\phi_\Zak^{h}(\alpha) = \varphi_\Zak^\alpha.
\end{equation}
Therefore the many-body Zak phase of the BI is related to the combined Zak phase of hole excitations from all sectors,
\begin{equation}
\phi_\Zak^{\rm BI} = \sum_{\alpha~\rm occ.} \phi_\Zak^{h}(\alpha).
\end{equation}

\section{Dimerized Mott insulators}
\label{sec:MIs}
In this section we show how the method for the detection of many-body Zak phases can be applied to interacting bosons in a one-dimensional (1D) lattice. In the hard-core limit the system is equivalent to non-interacting fermions by virtue of the Jordan-Wigner transformation. From the last section we know that in this limit the Zak phase of quasiparticle excitations is directly related to the many-body Zak phase of the groundstate. Here we study dimerized Mott insulators (MIs) which are generic examples for states with non-trivial many-body Zak phases \cite{Grusdt2013EdgeStates,GonzalezCuadra2019}.

\begin{figure}[t!]
\centering
\epsfig{file=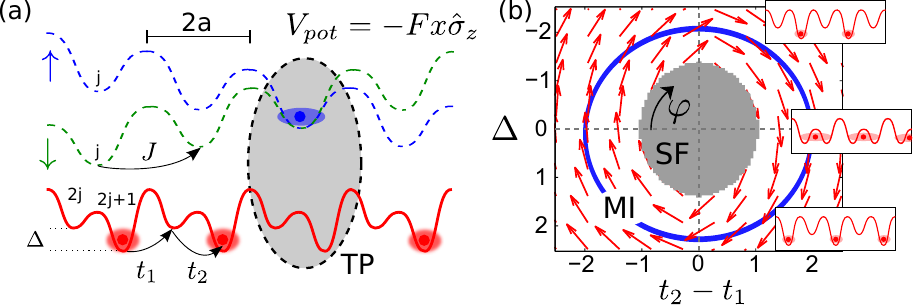, width=0.5\textwidth}
\caption{We consider a 1D model of interacting bosons in a superlattice potential at half filling, solid red in (a). For large enough $|\Delta|$ and $|t_2-t_1|$, a gapped Mott insulating (MI) phase is realized, while for small $|\Delta|$ or $|t_2-t_1|$ the system is superfluid (SF), see (b). By adiabatically changing $\Delta$ as well as $t_1$ and $t_2$ along a loop within the MI phase (parametrization $\varphi$, solid blue in (b)) a topological Thouless pump can be realized \cite{Hayward2018a} which is characterized by a many body Chern number. To measure this Chern number directly we couple a hole excitation of the MI to a two-component impurity in a conventional lattice, leading to the formation of a topological polaron (TP). Then, using a combination of Ramsey interferometry and Bloch oscillations of the impurity (driven by $V_{\text{pot}}$), the Zak phase of the TP can be measured. As indicated by arrows in (b), the winding of the Zak phase over parameter space yields the many-body Chern number. The perturbative results in (b) are given in units of $\min \l t_1,t_2 \r$.}
\label{fig:setup}
\end{figure}

In the calculations we include a mobile impurity interacting with the host many-body system. It binds to a quasiparticle excitation and forms a TP, see Fig.\ref{fig:setup} (a). By exact numerical simulations we demonstrate that the Zak phase of the TP is a direct measure for the many-body Zak phase of the MI.

\subsection{Model}
We consider the 1D superlattice Bose Hubbard model described by the following Hamiltonian,
\begin{multline}
\H_\text{B} = - \sum_{j} \l t_2 \bd_{2j+1} \b_{2j} +  t_1 \bd_{2j} \b_{2j-1} + \hc \r \\ 
+ \frac{\Delta}{2} \sum_{i} (-1)^i \bd_i \b_i + \frac{U}{2} \sum_i \bd_i \b_i \l \bd_i \b_i - 1 \r . 
\label{eq:Hdef}
\end{multline}
Here $t_{1,2}$ denote alternating hopping amplitudes, $\Delta$ is the strength of a staggered potential and $U$ is the interaction energy of two bosons (annihilation operator $\b_j$) occupying the same lattice site. 

We calculated the phase diagram of the model \eqref{eq:Hdef} at half filling using lowest order perturbation theory in Fig.\ref{fig:setup} (b). It consists of a gapless superfluid phase for small $U$, $\Delta$ or $|t_1-t_2|$ and a gapped Mott insulating phase (MI) otherwise. This system has recently been implemented experimentally using ultracold atoms \cite{Lohse2015}, and its phase diagram was studied more accurately by numerical DMRG simulations \cite{Hayward2018a}. 

In the MI phase the groundstate of Eq.\eqref{eq:Hdef} realizes a many-body topological Thouless pump \cite{Thouless1983,Berg2011,Lohse2015,Hayward2018a}. Changing the model parameters adiabatically in such a way that the superfluid phase is encircled leads to a quantized current. Trajectories through parameter space of this type can be described by the angle $\varphi$ as shown in Fig.\ref{fig:setup} (b). The current is directly proportional to the winding of the many-body Zak phase \cite{Xiao2010}, which defines the integer Chern number
\begin{equation}
 \mathcal{C} = \frac{1}{2 \pi} \int_{0}^{2 \pi} d\varphi ~ \partial_\varphi  \phi_\Zak(\varphi).
\label{eq:MBchernBulk}
\end{equation} 
When the system is inversion symmetric, realized for either $\Delta = 0$ or $t_1=t_2$ in Eq.\eqref{eq:Hdef}, the MI phase comes in two different symmetry-protected topological phases \cite{Zhu2013,Grusdt2013EdgeStates}. They can be distinguished by the quantized value of the many-body Zak phase $\phi_{\rm Zak}=0,\pi$ as will be explained in detail in Sec.\ref{sec:SPTexplanation} below. 

Now we introduce two-component impurities (annihilation operators $\c_{j,\sigma}$) which will be coupled to hole excitations of the MI. We place them into a long lattice with period $2a$, see Fig.\ref{fig:setup}. Their dynamics is described by
\begin{equation}
\H_\text{I} = - J \sum_{j,\sigma} \l \cd_{j,\sigma} \c_{j+1,\sigma}  + \hc \r - F \sum_{j, \tau,\sigma} 2 a j \cd_{j,\tau} \sigma^z_{\tau,\sigma} \c_{j,\sigma},
\label{eq:HImp}
\end{equation}
where $j$ labels the unit-cells and $\sigma, \tau = \uparrow, \downarrow$ are pseudospin indices. $J$ is the tunneling rate of the impurities in the long lattice. We also included external forces $\pm F$ acting differently on the two components. In experiments this could be realized by magnetic field gradients when the pseudospin components are realized by hyperfine sates \cite{Atala2012}. The case when the impurity lattice is replaced by a continous model can be discussed e.g. using the strong-coupling approximation of Ref.\cite{Grusdt2016TP}.

To achieve strong coupling between the impurity and the hole excitation we consider repulsive interactions between the impurity and the host bosons, which we describe by
\begin{equation}
\H_\text{IB} = V \sum_{j,\sigma} \cd_{j,\sigma} \c_{j,\sigma} \l \bd_{2j} \b_{2j} + \bd_{2j+1} \b_{2j+1}  \r.
\end{equation}
We take into account only the local interaction between an impurity at site $j$ and bosons in the two neighboring sites $i=2j$ and $i=2j+1$ of the boson lattice, see Fig.\ref{fig:setup}(a). When the system is realized using ultracold atoms in optical lattices with contact interactions, these terms are most relevant because the overlaps of the corresponding Wannier orbitals are maximal in this case.

\subsection{The protocol}
We start by describing the protocol for the measurement of the TP Zak phase in detail. The theoretical analysis follows in the subsequent sections. Here we discuss a specific situation relevant for ultracold atoms, but the basic ideas carry over to more general systems.

The first step consists of preparing the TP. If the impurity atoms ($\c_j$) are realized as long-lived electronic excited states of the groundstate bosons ($\b_i$), radio frequency pulses can be used to created a small concentration of TP wave packets with a given pseudospin (say $\uparrow$). We will assume that these TPs are initially at rest, i.e. their average momentum is $q=0$. The experimental feasibility of a similar preparation scheme has been demonstrated \cite{Weitenberg2011}. Note however that exact momentum resolution is not a necessary requirement. Next, applying a Ramsey $\pi/2$-pulse leads to a coherent superposition of $\uparrow$ and $\downarrow$ TPs and the single-TP state is described by
\begin{equation}
\ket{\Phi_\text{TP}(0)} = \l \ket{\uparrow, ~q=0} + \ket{\downarrow, ~0} \r / \sqrt{2}.
\end{equation}

Next we apply a magnetic field gradient which realizes the linear, spin-dependent potential (force $\pm F$) in Eq.\eqref{eq:HImp} when $\uparrow$, $\downarrow$ correspond to different hyperfine states with $m_F=\pm 1$ \cite{Atala2012}. This drives Bloch oscillations of the TP, and we keep the force switched on for one Bloch period $T_\text{B}$. Meanwhile each TP component crosses one Brillouin zone (BZ) and picks up a dynamical phase as well as a geometric Zak phase, 
\begin{equation}
\ket{\Phi_\text{TP}(T_\text{B})} = \frac{e^{i \phi_\text{dyn}}}{\sqrt{2}} \l \ket{\uparrow, -2 \pi}  e^{i \phi^\TP_\Zak}+ \ket{\downarrow, 2 \pi} e^{- i \phi^\TP_\Zak }\r.
\end{equation}

We note that this Zak phase $\phi^\TP_\Zak$ is a true many-body Zak phase, because the TP is a many-body excitation of the MI. A second $\pi/2$ Ramsey pulse can finally be used to read out the accumulated relative phase, i.e. $\Delta \phi = 2 \phi^\TP_\Zak$. Because each spin crossed the entire BZ we obtain twice the many-body Zak phase of the TP, while the dynamical phases of both components cancel. Notice that this result is true also if we start from TPs with a broad distribution of quasimomenta $q$.
 
When the system is inversion symmetric, it is sufficient to move every pseudospin component across half the BZ. The dynamical phases of both components are equal due to the symmetry $\omega_{\rm TP}(-k) = \omega_{\rm TP}(k)$ of the TP dispersion relation. In this case it is important to start from a sharp distribution of TP quasimomenta around $q=0$. The corresponding Ramsey signal equals the many-body Zak phase of the TP, $\Delta \phi = \phi^\TP_\Zak$. In systems without inversion symmetry, spin-echo techniques as suggested in Ref. \cite{Abanin2012} can also be applied to improve the protocol.
 
So far we have presented a measurement scheme for the many-body Zak phase $\phi_\Zak^\TP$ of the TP. The calculations in the next section confirm that it allows to measure the many-body Zak phase $\phi_\Zak$ of the host system. We will demonstrate that the inversion symmetry-protected many body Zak phase \cite{Grusdt2013EdgeStates} is equal to the TP Zak phase, which is also quantized by the inversion symmetry in this case. Moreover, the many-body Chern number $\mathcal{C}$ characterizing the Thouless pump, see Eq.\eqref{eq:MBchernBulk}, can be extracted from its TP counterpart $\mathcal{C}_{\rm TP}$. Here the Chern number of the TP is defined by the winding of its Zak phase 
\begin{equation}
\mathcal{C}_{\rm TP} = \frac{1}{2 \pi} \int_{0}^{2 \pi} d\varphi ~ \partial_\varphi  \phi_\Zak^{\rm TP}(\varphi)
\label{eq:MBchernTP}
\end{equation}
when the Hamiltonian is adiabatically changed, following the loop in parameter space described by the angle $\varphi$ in Fig.\ref{fig:setup} (b).

For the protocol to work, the TP should be close to the strong-coupling regime. As discussed in Ref.\cite{Grusdt2016TP} this allows to describe the TP using a product wavefunction,
\begin{equation}
\ket{\Phi_{\rm TP}(q,\sigma)} \approx \ket{\Phi_h(q)} \otimes \ket{\psi_{\rm I}(\sigma)},
\label{eq:defSC}
\end{equation}
where the impurity follows the motion of the hole adiabatically. Here $\ket{\Phi_h(q)}$ denotes the wavefunction of a hole excitation at quasimomentum $q$ while $\ket{\psi_{\rm I}}$ describes the wavefunction of the impurity bound to the hole. If the strong-coupling wavefunction \eqref{eq:defSC} applies, where $\ket{\psi_{\rm I}}$ is independent of $q$, it follows that the TP Zak phase is equivalent to the Zak phase of the hole.

Being in the strong-coupling regime requires the impurity to be sufficiently mobile, $J \gg t_{1,2}$ (see also Eq.\eqref{eq:defMobileImp} below). The impurity-boson interaction $V$ has to be sufficiently small not to open the bulk gap of the MI because this could destroy the topological phase completely around the impurity. On the other hand $V$ has to be sufficiently large to bind the impurity to the hole. Specific conditions for the model under consideration are discussed below.

 \subsection{Polaron transformation}
 \label{sec:LLP}
A powerful tool developed for the description of polarons in polarizable crystals is the Lee-Low-Pines (LLP) unitary transformation \cite{Lee1953}. It makes use of translational invariance and explicitly yields the total momentum $q$ as a conserved quantum number. The basic idea is to translate the entire Bose system by the impurity position,
\begin{equation}
\hat{U} = e^{i \hat{S}}, \quad \hat{S} = \sum_j 2ja \cd_j \c_j  \underbrace{\int_\text{BZ} dk ~ k \sum_{\alpha=1,2} \bd_{k,\alpha} \b_{k,\alpha}}_{=: \hat{P}_\text{B}}.
\label{eq:LLP}
\end{equation}
Here $\alpha=1,2$ is a band index and $\b_{k,\alpha}$ denotes the $k$-th Fourier component of the boson operators $\b_{2j}$ ($\alpha=1$) and $\b_{2j+1}$ ($\alpha=0$) respectively. For simplicity we suppress the spin label of the impurity from now on.

To calculate the Zak phase of TPs in our model, we apply the LLP transformation. To this end we simplify the impurity Hamiltonian first, $\H_\text{I} = - 2 J \int_\text{BZ} dq \cos(q - \omega_\text{B} t) \cd_q \c_q$. Here we introduced the Bloch oscillation frequency $\omega_\text{B} = 2 a F$ and eliminated the linear potential by performing a time-dependent gauge transformation. This is a standard trick to describe Bloch oscillations in finite systems with periodic boundary conditions. Note that this is also equivalent to imposing twisted boundary conditions \cite{Niu1985} for the impurity, with a time-dependent twist angle given by $\vartheta = \omega_\text{B} t$.

Applying the LLP transformation \eqref{eq:LLP} to the full Hamiltonian $\H=\H_\text{I} + \H_\text{B} + \H_\text{IB}$ with exactly one impurity yields
\begin{multline}
\tilde{\mathcal{H}} = \hat{U} \H \hat{U}^\dagger = \int_\text{BZ} dq ~ \cd_q \c_q \Bigl[ \H_\text{B} + V \l \bd_0 \b_0 + \bd_1 \b_1 \r  \\
- 2 J  \cos \l q - \omega_\text{B} t - \hat{P}_\text{B} \r  \Bigr] =:  \int_\text{BZ} dq ~ \cd_q \c_q \H_q(t).
\label{eq:PolaronHamiltonianImpCentered}
\end{multline}
This Hamiltonian is block diagonal in the total system momentum $q$, which can be changed in time by the driving force $F$. The groundstate of $\H_q$ is the TP state $\ket{\tilde{\Phi}_\TP(q)}$. Its energy $\omega_\TP(q)$ yields the TP dispersion relation, and the gap to the first excited state $\Delta_\TP(q)$ can be used as a measure for the TP binding energy. 

The Zak phase of the TP can be easily calculated in the LLP frame by adiabatically changing the total momentum $q$,
\begin{equation}
\phi_{\rm Zak}^{\TP} =  \int_{\rm BZ} dq~ \bra{\tilde{\Phi}_\TP(q)} i \partial_q \ket{\tilde{\Phi}_\TP(q)}.
\label{eq:defTPzakLLP}
\end{equation}
To avoid issues related to the freedom in choosing the global phase of $\ket{\tilde{\Phi}_\TP(q)}$ in the numerics we use a gauge-invariant discrete expression for the Berry phase, see e.g. Ref.\cite{Resta2007}.

Eq.\eqref{eq:defTPzakLLP} for the TP Zak phase after application of the LLP transformation is manifestly gauge invariant. Unlike in the case of the many-body Zak phase of the MI the definition of the unit-cell is fixed by the position of the impurity lattice. Hence the scheme does not suffer from fluctuations $\delta V$ of the overall potential $V = - F \hat{\sigma}^z x_{\rm I} +\delta V$ acting on the impurity (at position $x_{\rm I}$), which needed to be overcome in previous Zak phase measurements with a Bose-Einstein condensate \cite{Atala2012}.

\begin{figure}[t!]
\centering
\epsfig{file=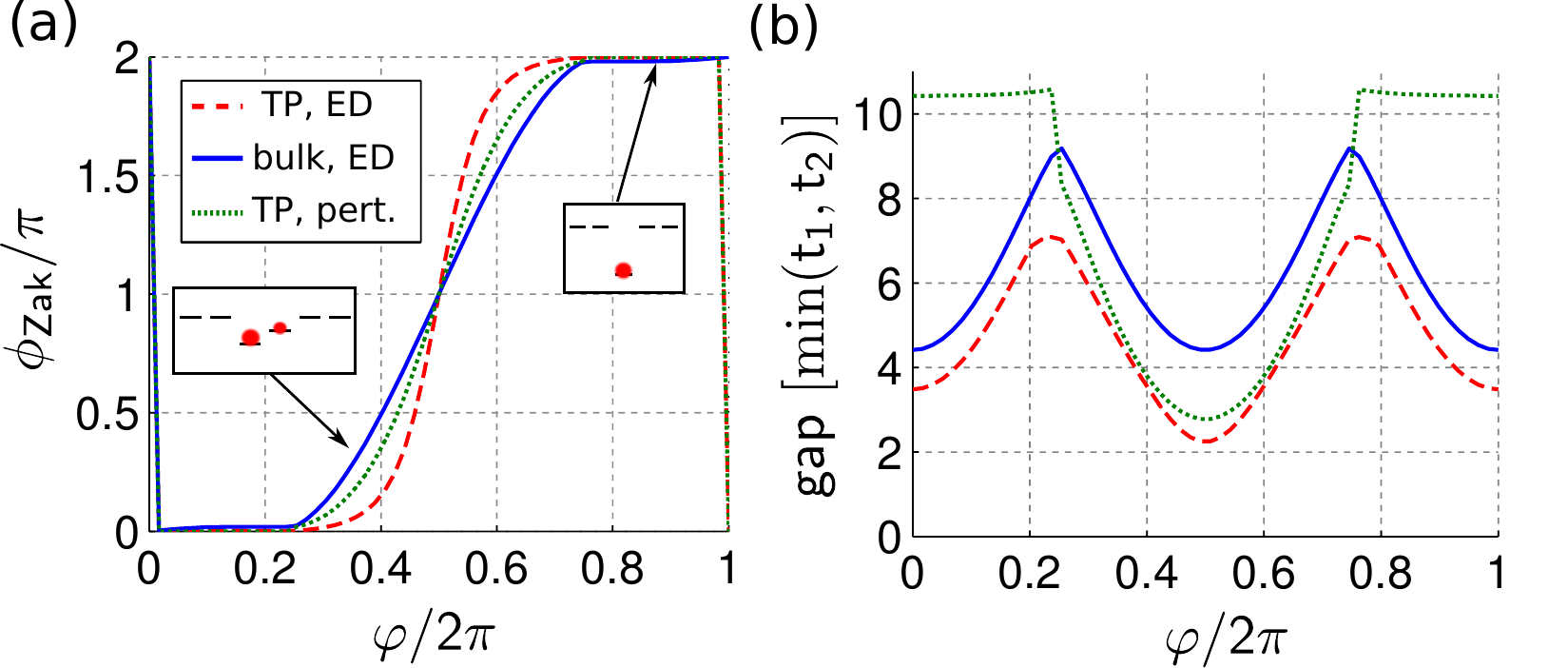, width=0.5\textwidth}
\caption{Numerical results from ED of \eqref{eq:PolaronHamiltonianImpCentered} along the Thouless pump cycle parametrized by $\varphi$. (a) Many-body Zak phases of the bulk (solid) for $5$ bosons and the TP (dashed) for $4$ bosons. For comparison the perturbative result for the TP is shown (dotted). Insets show the effective single hole models. (b) Gaps of the many-body groundstates, with same notations as in (a). We simulated $10$ sites and used $U=20$, $V=10$, $J=2$ and a radius $\sqrt{|t_2-t_1|^2+ \Delta^2}=5$ for the Thouless pump cycle shown in Fig.\ref{fig:setup} (b), all in units of $\min \l t_1,t_2 \r$.}
\label{fig:results}
\end{figure}

\subsection{Numerical results}
In Fig.\ref{fig:results} (a) we show the results obtained from exact diagonalization (ED) of $\H_q$ for realistic model parameters. We find that the TP Zak phase qualitatively follows the many-body Zak phase of the bulk as we change model parameters. Importantly the windings of both quantities over the Thouless pump cycle (i.e. the corresponding Chern numbers) coincide,
\begin{equation}
\mathcal{C}_\TP = \mathcal{C}.
\end{equation}

In Fig.\ref{fig:results} (b) we compare the TP gap and find that it is of the same order as the bulk gap of the MI. This is an important requirement to guarantee adiabaticity throughout an experiment. A simple perturbative analysis yields reasonable results, but we find large deviations when the bulk gap is too small. This is not surprising because in the perturbative analysis we did not take into account particle-hole fluctuations in the MI phase.

\begin{figure}[t]
\centering
\epsfig{file=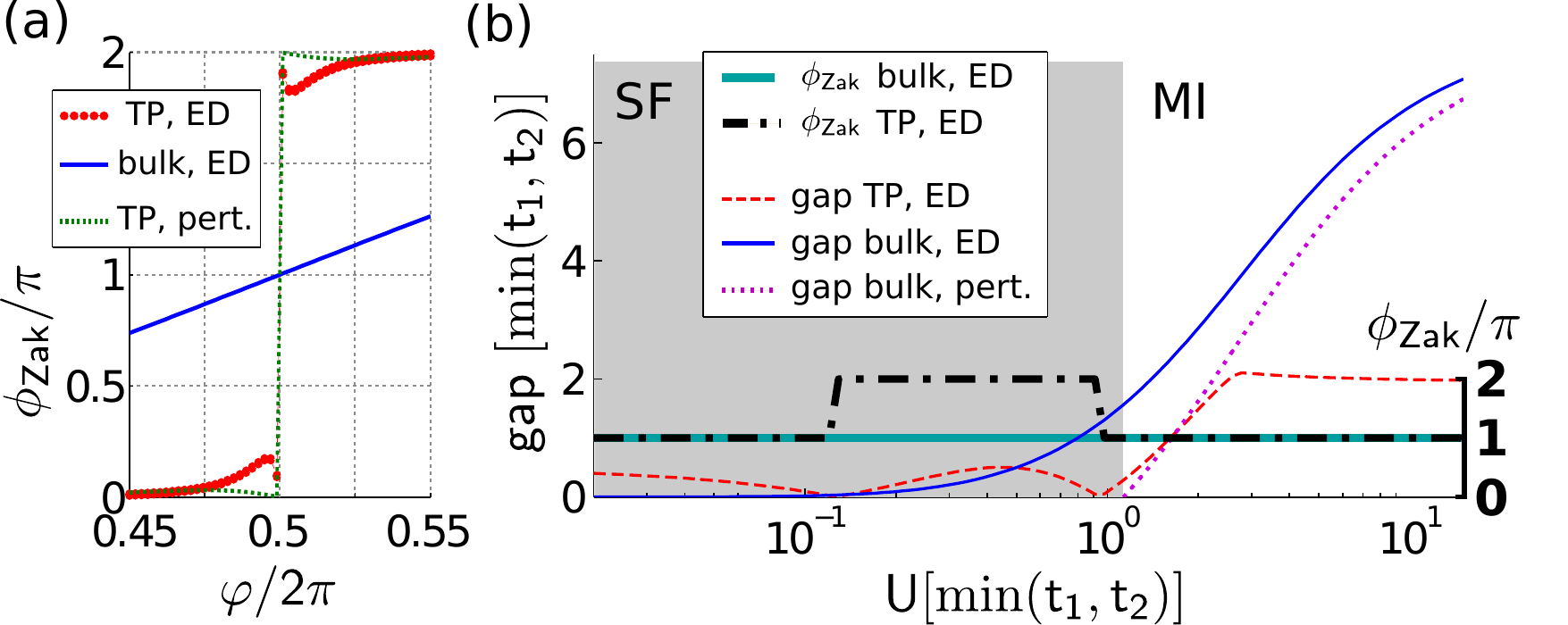, width=0.5\textwidth}
\caption{(a) The TP Zak phase of a quasi-static impurity (dotted line, $N=4$ bosons) deviates from the bulk many-body Zak phase (solid, $N=5$ bosons). We have chosen a small hopping $J=0.2 \min \l t_1,t_2 \r$ here and otherwise used the same parameters as in Fig.\ref{fig:results}. For comparison we show the perturbatively obtained TP Zak phase (dashed). (b) Many body Zak phases (thick lines) and excitation gaps (thin lines) are shown for a mobile impurity in an inversion-symmetric system. We used ED and varied the interaction $U$. Parameters were $V=10$, $J=2$ in units of $\min \l t_1,t_2 \r$, for $t_1=5 t_2$ and $\Delta=0$ and we used $10$ sites with $4$ bosons for the TP ($5$ bosons for bulk).}
\label{fig:results2}
\end{figure}

In Fig.\ref{fig:results2} (a) we show what happens in the case of a quasi-static impurity, where the TP is not in the strong-coupling regime. We have chosen a much smaller hopping $J=0.2 \min(t_1,t_2)$. In this case the TP Zak phase does \emph{not} wind when $\varphi$ is changed around the pumping cycle, unlike the many-body Zak phase of the MI. A more detailed discussion will be provided below, but essentially the impurity is not light enough to follow the hole excitation adiabatically. 

In Fig.\ref{fig:results2} (b) we investigate the interaction dependence of the TP Zak phase at inversion-invariant points $\Delta=0$. We observe that in the MI regime the TP Zak phase correctly reflects the bulk topological invariant $\phi_\Zak=\pi$. On the superfluid side on the other hand (we estimated the transition perturbatively), particle-hole fluctuations destroy the TP. The gap $\Delta_{\rm TP}$ closes and the TP Zak phase indicates a topological phase transitions. 

\subsection{Single hole approximation}
\label{sec:SingleHoleApprx}
To get a better understanding of the numerical results presented above, we will now restrict ourselves to only a single hole excitation of the MI. This is justified as long as no particle-hole pairs can be created in the vicinity of the mobile impurity. Furthermore we use a lowest order cell strong-coupling perturbative expansion technique \cite{Freericks1996,Buonsante2004} to describe the MI. Albeit simple, this method yields good results for the symmetry-protected version of our model \cite{Grusdt2013EdgeStates}. The key idea is to choose unit cells coupled by the smaller of the two couplings (chosen to be $t_2$ here), and start by solving the single particle problem of one bosons occupying each unit cell. The coupling $t_2$ can then be treated perturbatively. Using this technique we calculated the phase diagram and polarization vectors in Fig.\ref{fig:setup}(b). 

Using the same technique, we obtain the following effective Hamiltonian describing the dynamics of holes in the MI,
\begin{equation}
\H_\text{B} \approx -t_\text{h}  \sum_j  \l  \dd_{j+1} \d_{j} + \hc \r, \quad t_\text{h} = \frac{t_1 t_2}{\sqrt{4 t_1^2 + \Delta^2}}.
\label{eq:simplifiedHole}
\end{equation}
Here $\d_j$ annihilates a hole in unit cell $j$. To describe the binding to the impurity, we apply the LLP transformation to the simplified hole Hamiltonian \eqref{eq:simplifiedHole}. For a defect-free MI the total momentum vanishes, $\hat{P}_\text{B} \ket{\text{MI}}=0$, and we may write $\hat{P}_\text{B} = \int_\text{BZ} dk~ k ~ \dd_k \d_k$ in the subspace of small hole density.

When only a single hole is considered, we obtain the kinetic hole Hamiltonian after applying the LLP transformation
\begin{equation}
\H_q^{\text{kin}} = - \sum_j \Omega_q \dd_{j+1} \d_j + \hc,
\label{eq:EkinHoleLLP}
\end{equation}
where the effective nearest neighbor hopping is given by 
\begin{equation}
\Omega_q = t_\text{h} + J e^{i \l q  - \omega_\text{B} t \r}.
\label{eq:HoppingHoleLLP}
\end{equation}
The second term is a manifestation of the impurity kinetic energy in the LLP frame. The potential energy term will be included below.

To measure the Zak phase of the TP, a constant force $F$ is applied to the impurity for one Bloch cycle. In the LLP frame this gives rise to a force acting on the hole: As can be seen from Eq.\eqref{eq:HoppingHoleLLP} the complex phase of the hole hopping, ${\rm arg} ~\Omega_q$, changes in time. This corresponds to an artificial electric field which can give rise to the non-trivial Zak phase of the TP. Non-trivial Zak phases are only possible, however, if the complex phase of $\Omega_q$ changes adiabatically from $0$ to $2 \pi$ in one Bloch cycle, corresponding to a full twist of boundary conditions for the hole in the LLP frame. We thus require
\begin{equation}
t_\text{h} < J \qquad \text{(\emph{mobile impurity})},
\label{eq:defMobileImp}
\end{equation}
i.e. the TP has to be in the strong-coupling regime. For $t_\text{h} > J $ the impurity will be called quasi-static. In this case the heavy impurity can not follow the dynamics of the hole and it is thus not possible to map out the underlying topology of the MI.

To include the effect of boson-impurity interactions $V$ in the perturbative theory, we distinguish between a trivial case, when the hopping $t_2$ between interacting lattice sites $i=0,1$ in Eq.\eqref{eq:PolaronHamiltonianImpCentered} is the larger one ($t_2 > t_1$), and a non-trivial case when it is the smaller one ($t_2 < t_1$). In the trivial case, the effective hole potential reads 
\begin{equation}
\H_q^{\text{pot}} = -V \hat{n}_0
\end{equation}
where $\hat{n}_j=\dd_j \d_j$ denotes the hole density in unit cell $j$. In the non-trivial case in contrast, the hole at both $j=0,1$ is affected by the potential, 
\begin{equation}
\H_q^{\text{pot}} = -V_0 \hat{n}_0-V_1 \hat{n}_1.
\label{eq:nonTrivialHolePotLLP}
\end{equation}
These effective hole potentials are depicted in the inset of Fig.\ref{fig:results} (a). To zeroth order in $t_2$, they are given by
\begin{equation}
V_{0,1} = \frac{V}{2} \l 1 \pm \Delta / \sqrt{4 t_1^2 + \Delta^2} \r.
\end{equation}
This perturbative result is valid for $V\ll \sqrt{4 t_1^2 + \Delta^2}$. 

Using perturbation theory in the effective hole hopping $\Omega_q$ we can calculate the polarization in the groundstate of $\H_q^\text{pot}+\H_q^\text{kin}$. The hole is bound to the impurity which is localized in the center of the LLP frame. In the non-trivial case \eqref{eq:nonTrivialHolePotLLP} the bound state wavefunction is polarized due to the different potentials $V_0 \neq V_1$ acting on neighboring sites, see inset of Fig.\ref{fig:results} (a). On the other hand, in the trivial case the groundstate wavefunction is symmetric around the origin and the TP groundstate is unpolarized. 

The polarization of the hole wavefunction in the LLP frame gives rise to a geometric phase -- the TP Zak phase -- when the force $F$ is applied. We calculated this geometric phase using the perturbative TP wavefunction and compare with exact numerics in Fig.\ref{fig:results} (a). The simple perturbative analysis yields reasonable results for the Zak phases and predicts the correct winding as $\varphi$ changes continuously from $0$ to $2 \pi$.

\subsection{Symmetry-protected topological order}
\label{sec:SPTexplanation}
In Ref. \cite{Grusdt2013EdgeStates} it was pointed out that the half-filled super lattice Bose Hubbard model supports an (inversion-) symmetry-protected topological phase: For $\Delta=0$ the many-body Zak phase can only have two quantized values, 
\begin{equation}
\phi_\Zak=0,\pi.
\label{eq:quantizedZak}
\end{equation}

This quantization is a direct consequence of the inversion symmetry: If $\H(\vartheta)$ denotes the Hamiltonian for twisted boundary conditions with twist angle $\vartheta$, then 
\begin{equation}
\hat{I} \H(\vartheta) \hat{I} = \H(-\vartheta),
\end{equation}
where $\hat{I}$ is the inversion operator. Hence we obtain for the non-degenerate groundstate that $\ket{\psi(-\vartheta)} = e^{i \chi_\vartheta} \hat{I} \ket{\psi(\vartheta)}$, for a real function $\chi_\vartheta$. Using this result it follows for the Berry connection $\mathcal{A}(\vartheta) = \bra{\psi(\vartheta)} i \partial_\vartheta \ket{\psi(\vartheta)}$ that 
\begin{equation}
\mathcal{A}(-\vartheta) = - \mathcal{A}(\vartheta) + \partial_\vartheta \chi_\vartheta.
\end{equation}
Therefore we get $\phi_\Zak = \chi_\pi - \chi_0 \in \left\{ 0, \pi \right\}$. In the last step we used that $\hat{I} \ket{\psi(q)} = \pm \ket{\psi(q)}$ for $q=0,\pi$ such that $e^{i \chi_q}=\pm 1$ denotes the eigenvalue of the inversion operator at $q=0,\pi$ respectively.

The quantized Zak phase \eqref{eq:quantizedZak} allows to distinguish the two possible dimerizations of the MI, see Fig.\ref{fig:overview} (a). This becomes clear from an explicit calculation of the many-body Zak phase. To this end we consider the trivial limit where one of the hopping elements is zero, $t_2=0$ say. Here the groundstate is a product state with one boson per dimer. Introducing twisted periodic boundary conditions corresponds to using complex hopping elements whose phases sum up to $\vartheta$ when the system is encircled once. For simplicity we choose a gauge where only one of the hopping elements with amplitude $t_1$ is modified (say from site $i=1$ to $i=L$) and becomes $t_1 e^{i \vartheta}$. The dimer Hamiltonian for the boson corresponding to this bond now reads $- t_1 e^{i \vartheta} \bd_L \b_1 + \hc$ and its groundstate is
\begin{equation}
\ket{\psi_0(\vartheta)} = ( \bd_1 + e^{i \vartheta} \bd_L ) \ket{0}.
\label{eq:atomicDimer}
\end{equation}
Because this is the only dimer appearing in the product state which depends on $\vartheta$, we obtain the many-body Zak phase $\phi_{\rm Zak} = \pi$. If, on the other hand, the second dimer configuration is realized where $t_1=0$ and $t_2>0$, the twisted boundary conditions have no effect at all (because $t_1 e^{i \vartheta} = 0 \times e^{i \vartheta} = 0$) and thus $\phi_\Zak=0$. 

Because the setup including the impurity is inversion symmetric for $\Delta = 0$, the same arguments as above apply and it follows that the TP Zak phase is quantized as well
\begin{equation}
\phi_\Zak^\TP=0,\pi.
\end{equation}
This explains why TP and bulk Zak phases at  $\Delta=0$ (i.e. $\varphi=0,\pi$) in Fig.\ref{fig:results} (a) are strictly quantized. The simple calculations based on the single-hole approximation, see Sec.\ref{sec:SingleHoleApprx}, demonstrate again that the TP determines the dimerization pattern of the MI. We conclude that the protocol can be used to detect symmetry-protected topological invariants, as long as the system is in the MI phase, see Fig.\ref{fig:results2}(b).

\section{Other systems with topological order}
\label{sec:OtherSystems}
Now we turn our attention to more general many-body systems in one dimension which have (symmetry-protected) topological order, see Fig. \ref{fig:overview} (b). We start by considering exactly solvable toy models for which we demonstrate that their many-body Zak phases can be measured by the use of topological polarons. Specifically we discuss topological superconductors supporting Majorana edge states (Sec.\ref{sec:TopSuperconductors}) and anti-ferromagnetic chains of spin $S=1/2$ and $S=1$ particles (Sec.\ref{sec:SpinChains}). Away from the analytically tractable points of these models we use exact numerical diagonalization to calculate the properties of topological polarons. 

The goal of this section is to demonstrate that the topological polaron concept can be used to detect the topological order in generic one-dimensional systems. The development of realistic schemes to implement the topological polarons experimentally will be devoted to future work.

\subsection{Topological superconductors}
\label{sec:TopSuperconductors}
Read and Green \cite{Read1999} have shown that fully gapped $p_x-ip_y$ spin-polarized superconducting states can be constructed in two dimensions which are characterized by a non-zero Chern number of their Bogoliubov quasiparticle excitations. Following Ref.\cite{Grusdt2016TP} this Chern number of the Bogoliubov quasiparticles could be measured by coupling them to mobile impurities serving as coherent probes of the many-body system.

Here we consider a simpler scenario and study $p$-wave superconductors in one dimension. Kitaev \cite{Kitaev2001} has generalized Read and Green's scenario to one-dimensional chains and predicted a topological phase transition from a trivial strong-pairing phase to a non-trivial weak-pairing phase. His second prediction that isolated Majorana fermions are localized at the edges of the chain in the weak-coupling phase has raised considerable interest because of their potential importance for robust quantum information processing. Following theoretical proposals \cite{Lutchyn2010,Oreg2010,Nadj-Perge2013} first experimental signatures for Majorana fermions have been found in different systems \cite{Mourik2012,NadjPerge2014}. However an unambiguous proof of the topological nature of these systems is still lacking.

\subsubsection{Kitaev chain, its excitations, and their many-body Zak phase}
\label{sec:KitaevChain}
The Kitaev model \cite{Kitaev2001} describes a chain of fermions $\a_j$ by the following BCS mean-field Hamiltonian,
\begin{multline}
\H = \sum_j \biggl[ -w \l \ad_j \a_{j+1} + \ad_{j+1} \a_j \r - \mu \l \ad_j \a_j - 1/2 \r \\
+ \Delta \a_j \a_{j+1} + \Delta^* \ad_{j+1} \ad_j  \biggr].
\label{eq:Kitaev}
\end{multline}
Assuming periodic boundary conditions, the Hamiltonian \eqref{eq:Kitaev} can be solved by a Bogoliubov transformation. As a result one obtains $\H = - \sum_k \bd_k \b_k \omega_k$ where $\omega_k = [(2 w \cos k + \mu)^2 + 4 |\Delta|^2 \sin^2 k]^{1/2}$  \cite{Kitaev2001}. The new fermions $\b_k = u_k \ad_{-k} + v_k \a_k$ in momentum space are related to the original fermions $\a_k$ and $\ad_k$ by a Bogoliubov transformation. In terms of the new fermions $\b_k$ the groundstate of the Kitaev chain $\ket{\psi_{\rm K}} = \prod_k \bd_k \ket{0}$ corresponds to a band insulator, see Sec.\ref{sec:BIs}.

The topological order of the Kitaev chain is determined by the Bloch vector $\ket{u(k)} = (u_k,v_k)^T$ which appears in the Bogoliubov transformation. It is an eigenvector of the Hamiltonian $h_k = \left[ 2 w \cos (k) + \mu \right] \sigma_z + \Delta \sin (k) \sigma_y$, where we assumed that $\Delta$ is real. Because of its particle-hole symmetry, $\{ h_k , \sigma_x \}=0$, the corresponding Zak phase is quantized, see Refs.~\cite{Ryu2002,Hatsugai2006},
\begin{equation}
\phi_{\rm Zak}^{\rm K} = \int_{\rm BZ} dk~ \bra{u(k)} i \partial_k \ket{u(k)} = 0, \pi.
\end{equation}
An alternative classification of the two resulting topological phases was introduced by Kitaev \cite{Kitaev2001}, who introduced the number of unpaired Majorana fermions in a system with open boundaries as a topological invariant. It was shown explicitly in Ref.~\cite{Budich2013a} that the two definitions are equivalent. 

The Bogoliubov quasiparticle excitations of the Kitaev chain correspond to hole excitation of the new fermions $\b_k$,
\begin{equation}
\ket{\Phi_{\rm B}(k)} = \b_k \ket{\psi_{\rm K}}.
\end{equation}
From the discussion of band insulators in Sec.\ref{sec:BIs} it follows that their Zak phase is identical to the Zak phase of the Kitaev chain, $\phi_{\rm Zak}^{\rm B}=\phi_{\rm Zak}^{\rm K}$.

\subsubsection{Topological polarons in the Kitaev chain}
\label{sec:KitaevChainTPs}
To bind a mobile impurity to a quasiparticle excitation in the Kitaev chain and form a TP, an adequate interaction between the impurity and the underlying fermions has to be realized. A natural choice would be a simple point interaction, which gives rise to Shiba states when the impurity is localized \cite{Shiba1968,Sau2013}. However, because the emerging Bogoliubov quasiparticles are a superposition of particle and hole states, the Shiba state of a mobile impurity acquires a non-trivial spatial structure. As we show by analytic and numerical calculations in Appendix \ref{sec:TopShibaState}, this leads to an additional contribution to the TP Zak phase and complicates a direct measurement of the quasiparticle Zak phase.

Here we choose an alternative route and construct an impurity-fermion interaction which leads to a simple TP bound state. To this end we consider the limits $w=\Delta=0$, $\mu \neq 0$ (topologically trivial, $\phi_\Zak^{\rm K}=0$) and $w = \Delta>0$, $\mu=0$ (non-trivial, $\phi_\Zak^{\rm K}=\pi$) discussed by Kitaev. In this case the eigenstates can most easily be constructed by decomposing every fermion into a pair of Majoranas,
\begin{equation}
\g_{2j-1} = \ad_j + \a_j, \qquad \g_{2j} = i ( \ad_j - \a_j ),
\end{equation}
where $j$ labels lattice sites.

In the topologically trivial phase Majorana fermions belonging to the same original fermions are paired, $\H_{\rm K} = \mu \sum_j \ad_j \a_j$ and the groundstate reads $\ket{\psi_{\rm BI}}=\prod_j \ad_j \ket{0}$. In the non-trivial phase, on the other hand, Majorana fermions belonging to neighboring fermions are paired, $\H_{\rm K} = 2 w \sum_j \tilde{a}^\dagger_j \tilde{a}_j$ with
\begin{equation}
\tilde{a}_j = \frac{1}{2} ( \g_{2j} + i \g_{2j+1} ), \quad \tilde{a}_j^\dagger = \frac{1}{2} ( \g_{2j} - i \g_{2j+1} ).
\label{eq:aTildeDef}
\end{equation}
The superconducting groundstate is the vacuum of these new fermions, $\tilde{a}_j \ket{\psi_{\rm SC}} = 0$. 

To construct TPs we consider a two-component mobile impurity $\c_{j,\sigma}$ described by the Hamiltonian $\H_{\rm I}$ in Eq.\eqref{eq:HImp}. It can tunnel between neighboring sites of a lattice (hopping amplitude $J$) which has the same orientation as the Kitaev chain \eqref{eq:Kitaev}. To measure the TP Zak phase interferometrically, a force is included which has opposite signs for the two internal states $\sigma=\uparrow,\downarrow$ of the impurity.

\begin{figure}[b]
\centering
\epsfig{file=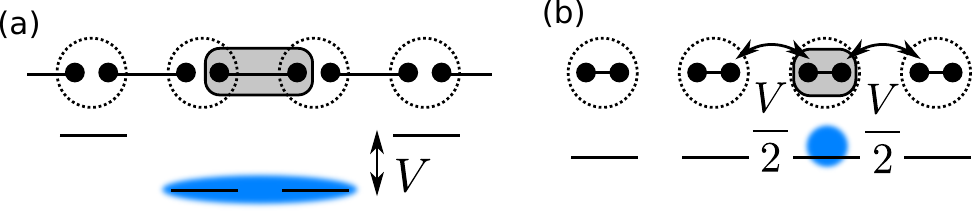, width=0.45\textwidth}
\caption{The TP wavefunction of an impurity (blue) bound to a Bogoliubov quasiparticle (gray) is shown. Black dots correspond to Majorana fermions, constructed from the original fermions (dashed circles). In (a) the topologically non-trivial phase is shown, where $\phi_\Zak^\TP=\pi$. In (b) the trivial case is shown, where $\phi_\Zak^\TP=0$.}
\label{fig:KitaevChainTP}
\end{figure}

\begin{figure*}[t!]
\centering
\epsfig{file=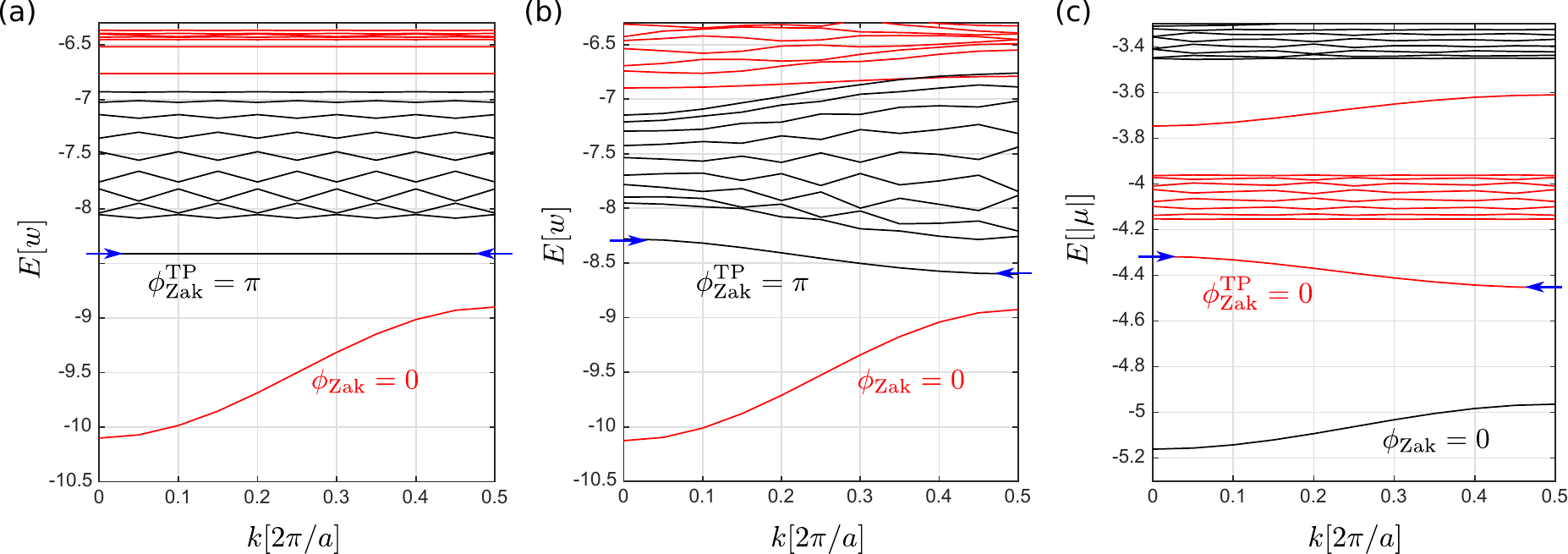, width=0.93\textwidth}
\caption{The TP band structure is calculated for the impurity-fermion interaction described by Eq.\eqref{eq:HIForiginal} using exact diagonalization, for a system of length $L=10$ with periodic boundary conditions. The fermion parity $P$, see Eq.\eqref{eq:defFParity}, is indicated by the color code (red: $P=1$, black: $P=0$). The TP state where the impurity binds to a single Bogoliubov quasiparticle excitation is marked by blue arrows. In the topologically non-trivial phase (a), for $w=\Delta$, $\mu=0$, the TP is localized (flat band) and its Zak phase is non-trivial as expected. We set $J=0.3 w$ and $V=0.5 w$ in the simulation. In (b) we included a chemical potential, $\mu=0.2 w$. As a result the TP band becomes dispersive, but the topology is the same as in (a). In the topologically trivial case (c), for $\mu=-1$, $\Delta=w=0$, the Zak phase of the TP is trivial. We have set $J=0.05 |\mu|$ and $V=0.5 |\mu|$ in this case.}
\label{fig:KitaevTPs}
\end{figure*}

\paragraph{Non-trivial phase.--}
In the topologically non-trivial phase, a Bogoliubov excitation around $j=0$ is described by $\ket{\Phi} = \tilde{a}^\dagger_0 \ket{\psi_{\rm SC}}$. Importantly, this excitation is localized on the bond between site $j=0$ and $j=1$, see Eq.\eqref{eq:aTildeDef}. 

To bind the impurity to the quasiparticle we consider the following interaction with the emerging fermions,
\begin{equation}
\H_{\rm IF} = -V \sum_{j,\sigma} \cd_{j,\sigma} \c_{j,\sigma} \left[ \tilde{a}^\dagger_{j-1} \tilde{a}_{j-1} + \tilde{a}^\dagger_{j} \tilde{a}_{j} \right].
\label{eq:defHIF}
\end{equation}
For the state $\ket{\Phi} = \tilde{a}^\dagger_0 \ket{\psi_{\rm SC}}$ the effective Hamiltonian of the impurity thus reads
\begin{equation}
\H = -J \sum_{j,\sigma} [ \cd_{j+1,\sigma} \c_{j,\sigma} + \hc ] - V \sum_\sigma [ \cd_{0,\sigma} \c_{0,\sigma} + \cd_{1,\sigma} \c_{1,\sigma}] 
\end{equation}
in the absence of the force, see Fig.\ref{fig:KitaevChainTP} (a). In the groundstate for $V \gg J$ the impurity is localized on the link between sites $j=0$ and $j=1$ and forms a dimer. I.e. (an immobile) TP is formed, described by the wavefunction
\begin{equation}
\ket{\Phi_\TP} = \frac{1}{\sqrt{2}} \l \cd_0 + \cd_1 \r \ket{0} \otimes \tilde{a}^\dagger_0 \ket{\psi_{\rm SC}},
\end{equation}
where we neglected the spin-index $\sigma$ of the impurity for simplicity.

When a weak force $F$ acting on the impurity is switched on for one Bloch period, the impurity picks up the geometric Zak phase which is quantized to values 
\begin{equation}
\phi_\Zak^{\TP}=0,\pi
\label{eq:TPZakQuant}
\end{equation}
due to inversion symmetry. Similar to the non-trivial case discussed in Sec.\ref{sec:SingleHoleApprx} a perturbative calculation in $J/V\ll 1$ shows that the TP Zak phase is non-trivial in this case, $\phi_\Zak^{\TP}=\pi$. As expected, the TP measures the Zak phase of the Bogoliubov quasiparticle, $\phi_{\rm Zak}^{\rm K}=\pi$.

\paragraph{Trivial phase.--}
In the topologically trivial phase, a Bogoliubov excitation localized at site $j=0$ is described by $\ket{\Phi} = \a_0 \ket{\psi_{\rm BI}}$. In this case, too, the Hamiltonian Eq.\eqref{eq:defHIF} allows to bind the impurity to the quasiparticle and form a TP. To see this, we express \eqref{eq:defHIF} in terms of the original fermions,
\begin{multline}
\H_{\rm IF} = \frac{V}{2} \sum_{j,\sigma} \cd_{j,\sigma} \c_{j,\sigma} \biggl[ \ad_{j+1} \a_j + \ad_j \a_{j-1} -\\
 - \ad_{j+1} \ad_j - \ad_{j} \ad_{j-1} + \hc \biggr].
 \label{eq:HIForiginal}
\end{multline}

In the following discussion of TPs we consider the case when $w \gg J,V$ which allows to discard the pairing terms in \eqref{eq:HIForiginal}. Moreover we consider the limit $V \gg J$, where the impurity is tightly bound to the quasiparticle. First we neglect the impurity hopping all together and obtain the lowest order groundstates (see Fig.\ref{fig:KitaevChainTP} (b))
\begin{equation}
\ket{\Phi_\TP(j)} = \cd_j \ket{0} \otimes \frac{1}{2} \l  \a_{j-1} -  \sqrt{2} \a_j + \a_{j+1}\r \ket{\psi_{\rm BI}}.
\end{equation}
We include small $J \ll V$ by using degenerate perturbation theory and obtain the following effective TP Hamiltonian,
\begin{equation}
\H_\TP^{\rm eff} = \frac{J}{\sqrt{2}} \sum_j \ket{\Phi_\TP(j+1)}\bra{\Phi_\TP(j)} + \hc.
\label{eq:HeffTPtrivial}
\end{equation}

When the weak force $F$ acting on the impurity is switched on for one Bloch period, the TP picks up the Zak phase. Because \eqref{eq:HeffTPtrivial} describes a pure nearest neighbor hopping Hamiltonian, the TP Zak phase is trivial, $\phi_\Zak^{\TP}=0$. It coincides with the Zak phase of the Bogoliubov quasiparticle, $\phi_{\rm Zak}^{\rm K}=0$.

In summary, we have shown for Kitaev chains with immobile Bogoliubov quasiparticles that TPs can be formed whose Zak phase allows a direct measurement of the quasiparticle topology,
\begin{equation}
\phi_\Zak^{\TP}=\phi_\Zak^{\rm K}.
\end{equation}
This represents a direct probe of the topological order in the system. Because the TP Zak phase is quantized, see Eq.\eqref{eq:TPZakQuant}, the result also holds more generally even when the quasiparticles become mobile or the couplings in the Hamiltonian are changed, as long as the system does not undergo a phase transition. We demonstrate this now by exact numerical calculations.

\subsubsection{Numerical simulations}
\label{sec:KitaevChainTPsNumerics}
In Fig.\ref{fig:KitaevTPs} we show the TP band structure, calculated for the impurity-fermion interaction $\H_{\rm IF}$ defined in Eq.\eqref{eq:HIForiginal}. We used the same numerical techniques as presented in Sec.\ref{sec:LLP} and extracted the Zak phase of the interacting mobile impurity. 

The lowest band corresponds to an unbound impurity propagating through the ground state of the Kitaev chain, with a band-width of $4 J$. The Zak phase is trivial in this case, $\phi_{\rm Zak}=0$, independent of the parameters for the Kitaev chain. 

The first excited state has one Bogoliubov excitation, and hence a different fermion number parity
\begin{equation}
\hat{P}  = \sum_j \cd_j \c_j \mod 2
\label{eq:defFParity}
\end{equation}
from the ground state. Therefore the first band is stable and can not decay into the ground state without violating parity conservation. For every total conserved momentum $k$ there are $L/a$ states, corresponding to the number of quasiparticle positions for a given impurity configuration. The lowest of these states corresponds to the TP bound state which we are interested in. It is protected by a gap of order $\sim V$ from the remaining $L/a-1$ states in the scattering continuum.

In Fig.\ref{fig:KitaevTPs} (a) the topologically non-trivial case discussed in the text is shown. As expected, we find a flat band corresponding to a localized TP, with a non-trivial Zak phase $\phi_{\rm Zak}^{\rm TP}=\pi$. In Fig.\ref{fig:KitaevTPs} (b) we included a finite chemical potential, such that the Bogoliubov quasiparticles become dispersive. As a result also the TP band has a finite dispersion, while its topological properties remain unchanged. In Fig.\ref{fig:KitaevTPs} (c) the topologically non-trivial case is shown. Although the Bogoliubov quasiparticles are localized in this case, the TP is dispersive as predicted in Eq.\eqref{eq:HeffTPtrivial}. Its Zak phase is trivial, $\phi_{\rm Zak}^{\rm TP}=0$, and provides a direct measure of the topology in the Kitaev chain.

\subsection{Spin chains with topological order}
\label{sec:SpinChains}
Now we turn our attention to gapped anti-ferromagnetic (AFM) spin chains in one dimension. Haldane conjectured that the groundstate of the integer spin, $S =1,2,...$, Heisenberg model is gapped \cite{Haldane1983b}. For the simplest case, $S=1$, degenerate edge states were found on the ends of systems with open boundary conditions \cite{Kennedy1990}, each corresponding to a spin $1/2$. These provide a clear signature for the existence of topological order in this model. Indeed the Haldane $S=1$ phase can be distinguished from other $S=1$ phases without degenerate edge states by a non-local string order parameter \cite{denNijs1989,Kennedy1992,Endres2011}. Here, instead, we use twisted periodic boundary conditions to quantify and detect the topological order of gapped spin models.

Important insight in the topological order of AFM spin chains can be obtained by employing the valence bond picture. It is based on the observation that the groundstate of two anti-ferromagnetically coupled spin $1/2$ particles,
\begin{equation}
\H = J \left[ \hat{S}^z_1 \hat{S}_2^z + e^{i \vartheta} \hat{S}_1^- \hat{S}_2^+ + e^{- i \vartheta} \hat{S}_1^+ \hat{S}_2^- \right]
\end{equation}
with an arbitrary phase $\vartheta$, is a singlet state
\begin{equation}
\ket{{\rm VB}(\vartheta)} = \l \ket{\uparrow \downarrow} - e^{i \vartheta} \ket{\downarrow \uparrow} \r / \sqrt{2}
\label{eq:twistedVBstate}
\end{equation}
also referred to as a valence bond (VB) state. 

Like the atomic dimer states discussed above, see Eq.\eqref{eq:atomicDimer}, the VB state has a non-trivial Berry (or Zak) phase 
\begin{equation}
\phi_\Zak^{\rm VB} = \int_0^{2 \pi} d\vartheta ~ \bra{{\rm VB}(\vartheta)} i \partial_{\vartheta} \ket{{\rm VB}(\vartheta)}  = \pi
\end{equation}
when $\vartheta$ is varied adiabatically by $2 \pi$. Hatsugai \cite{Hatsugai2006,Hatsugai2007} has shown that in this case the Zak phase is quantized e.g. by time-reversal symmetry and he suggested that it can be used as a local topological order parameter for gapped spin chains. Similar to the cases of topological superconductors and dimerized Mott insulators discussed previously, topologically inequivalent states differ by their VB pattern, see illustration in Fig.\ref{fig:overview} (a). Different patterns reflect themselves in the pattern of local Berry phases \cite{Hatsugai2006}. An equivalent theory based on twisted periodic boundary conditions is summarized in Appendix \ref{sec:TwistedPBCSpinChain}.

One of the simplest spin models with topological order is the VB solid constructed by Majumdar and Gosh \cite{Majumdar1969a,Majumdar1969}. They provided an exact solution, in terms of VB states, of a spin $1/2$ AFM. We discuss this model in detail below (\ref{sec:MajumdarGoshModel}) and show how TPs can be used to distinguish the two topologically inequivalent groundstates. Affleck et al. \cite{Affleck1987} constructed an exactly solvable Hamiltonian (the AKLT model) closely related to the $S=1$ AFM Heisenberg model and showed that its groundstate wavefunction can be understood as a VB solid which, however, does not break the translational symmetry. It is understood nowadays that the groundstates of the AKLT model and the AFM $S=1$ Heisenberg model are in the same topological class \cite{Pollmann2012}. The AKLT model is briefly discussed in Appendix \ref{sec:HaldaneSpinModel}.

\subsubsection{Majumdar-Gosh model}
\label{sec:MajumdarGoshModel}
The Majumdar-Gosh model \cite{Majumdar1969a} is defined by the following AFM spin-$1/2$ Hamiltonian,
\begin{equation}
\H_{\rm MG} = \frac{t}{2} \sum_{j=1}^N \hat{\vec{S}}_j \cdot  \hat{\vec{S}}_{j+1} + \frac{t}{4} \sum_{j=1}^N \hat{\vec{S}}_j \cdot  \hat{\vec{S}}_{j+2}
\end{equation}
with periodic boundary conditions (we assume $N$ is even for now). In its groundstate every three subsequent spins couple to the minimum possible total spin, $\hat{\vec{L}}_j^2 := ( \hat{\vec{S}}_{j}+\hat{\vec{S}}_{j+1}+\hat{\vec{S}}_{j+2})^2=3/4$. Majumdar has shown that there exist exactly two VB configurations with this property, which are separated by a finite gap from all other eigenstates \cite{Majumdar1969}. In the first case $\hat{\vec{S}}_{2l}$ and $\hat{\vec{S}}_{2l-1}$ form a VB for all $l=1...N/2$, whereas in the second case $\hat{\vec{S}}_{2l}$ and $\hat{\vec{S}}_{2l+1}$ are paired in a VB state. I.e. the two states correspond to different VB configurations as illustrated in Fig.\ref{fig:overview} (a). They are topologically distinct and give rise to different spin-Zak phases $\phi_\Zak^S=0$ and $\phi_\Zak^S=\pi$ (see definition in Appendix \ref{sec:TwistedPBCSpinChain}), quantized by the inversion symmetry in this case.

\begin{figure}[t]
\centering
\epsfig{file=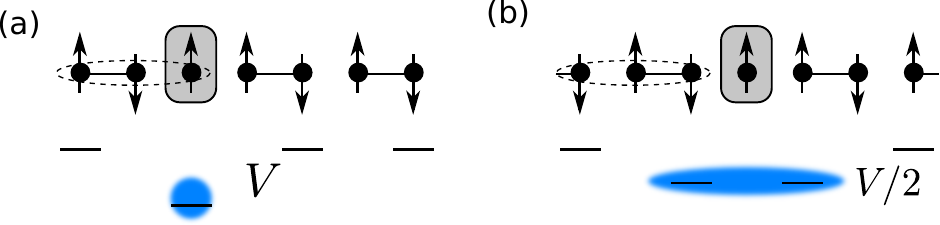, width=0.45\textwidth}
\caption{The TP wavefunction of an impurity (blue) bound to a topological excitation (domain wall) of the Majumdar-Gosh chain is shown. In the groundstate of the Majumdar-Gosh Hamiltonian every three subsequent spins (indicated by dashed circles) couple to total spin $1/2$. In (a) the topologically trivial phase is shown, where $\phi_\Zak^{\TP,\rm a}=0$. In (b) the non-trivial case is shown, where $\phi_\Zak^{\TP, \rm b}=\pi$.}
\label{fig:SpinChainTP2}
\end{figure}

\paragraph{Topological excitations.--}
To investigate the topological order in the Majumdar-Gosh model, we consider its topological excitations which correspond to domain walls between two different VB configurations, see Fig.\ref{fig:SpinChainTP2}. The first type of quasiparticle (a) is located between a bulk VB solid with $\phi_\Zak^S=0$ to the left and $\phi_\Zak^S=\pi$ to the right. For the second type of quasiparticle (b), the other way around, the VB solid to the left has $\phi_\Zak^S=\pi$ and the one to the right has $\phi_\Zak^S=0$. Note that unlike the overall values of the spin-Zak phases their differences are independent of the gauge choice and can be measured.

The spin-Zak phases of the two types of quasiparticle excitations differ by
\begin{equation}
\phi_{\rm Zak}^{S, \rm b} - \phi_{\rm Zak}^{S, \rm a} =\pi
\label{eq:DeltaPhiZakS}
\end{equation}
and thus allow to distinguish between the two topologically inequivalent configurations. To show this we note that one type of quasiparticle can be transformed into another by a global translation of the system by one lattice site. This changes the spin-Zak phase by $\pi$, as a consequence of the (generalized) King-Smith and Vanderbilt theorem \cite{Kingsmith1993,Resta1998} stating that the polarization $P$ in units of the lattice constant $d$ is directly related to the Zak phase, $P=d \phi_{\rm Zak}/2 \pi$. Formulated in terms of spins it reads
\begin{equation}
\sum_j  j \langle S_j^z \rangle / N = \phi_{\Zak}^S/2 \pi,
\end{equation}
as can be shown directly from Eq.\eqref{eq:UthetaSpinZak} in the appendix. Using the same techniques as in Sec.\ref{subsec:MBzakAndTwistedBCs} it follows for a single spin ($S_z^{\rm tot}=1/2$) in a Bloch band consisting of $N$ states that $\Delta \phi_\Zak^S = \pi$ for a translation by half a unit cell.

\paragraph{Topological polarons.--}
To construct TPs which allow a direct measurement of the quasiparticle spin-Zak phases, we consider an impurity hopping on a lattice with twice the period $a$ of the host spin system. Because the translational invariance is explicitly broken by the Majumdar-Gosh VB solid, this defines the true size of the unit cell of the system. Besides the free impurity Hamiltonian $\H_{\rm I}$ from Eq.\eqref{eq:HImp} we consider an impurity-spin interaction allowing to bind the impurity to the topological excitation and form a TP. The basic strategy is to couple to the excess spin $1/2$ of the quasiparticle excitation, see Fig.\ref{fig:SpinChainTP2}. 

Here we construct an interaction Hamiltonian in the strong coupling regime in which the quasiparticle tunneling (of the order $t$) is much smaller than the impurity energies, $t \ll J, V$. While this simplifies the analysis considerably and allows us to derive the TP groundstate and its Zak phase perturbatively, it should not be considered as a necessary requirement for realizing TPs in spin systems. 

We consider the following coupling,
\begin{equation}
\H_{\rm IS} = -V \sum_{j,\sigma} \cd_{j,\sigma} \c_{j,\sigma} \left[ \frac{1}{2} \hat{\vec{L}}_{j-2}^2 + \hat{\vec{L}}_{j-1}^2  + \frac{1}{2} \hat{\vec{L}}_{j}^2 \right],
\label{eq:defHIS}
\end{equation}
which commutes with the Majumdar-Gosh Hamiltonian, $[\H_{\rm IS},\H_{\rm MG}]=0$. For $V \gg J$ it binds the impurity to the topological defect as sketched in Fig.\ref{fig:SpinChainTP2}. Because $t \ll J$ we can treat the quasiparticle tunneling perturbatively and solve for the bound state of the impurity in the TP for a stationary topological excitation. For (a)-type quasiparticles, the resulting potential seen by the impurity leads to a tightly bound TP with a trivial Zak phase $\phi_{\Zak}^{\TP, \rm a}=0$, see Fig.\ref{fig:SpinChainTP2} (a). For (b)-type quasiparticles on the other hand the impurity remains delocalized over the two lattice sites around the domain wall and thus the corresponding TP Zak phase is non-trivial, $\phi_\Zak^{\TP,\rm b}=\pi$. As expected from Eq.\eqref{eq:DeltaPhiZakS} the difference of the two TP Zak phases is
\begin{equation}
\phi_{\rm Zak}^{\TP, \rm b} - \phi_{\rm Zak}^{\TP, \rm a} =\pi.
\end{equation}

\section{Summary and Outlook}
\label{sec:SummaryOutlook}
In summary, we discussed various gapped models of interacting particles in one dimension which have symmetry-protected topological order described by the many-body Zak phase. This includes band insulators, dimerized Mott insulators, topological superconductors and anti-ferromagnetic spin chains. We introduced a scheme for the direct measurement of their many-body Zak phases, based on two key ideas. First, an identification of the many-body Zak phase of the bulk system with the Zak phase corresponding to elementary excitations. We established such relations for all the models mentioned above. Second, the binding of a mobile impurity to the elementary excitation, which serves as a coherent probe of the many-body system. In this step a new quasiparticle is formed, the topological polaron \cite{Grusdt2016TP}. By applying different forces to the internal states of the impurity, Ramsey interferometry allows to measure the Zak phase of the topological polaron. We have shown by explicit calculations that the topological order of the host many-body system can be mapped out in this way. Notably there is no need to realize twisted periodic boundary conditions experimentally, which are used as a theoretical tool to define many-body Zak phases in the first place.  

As a concrete experimentally relevant example, we considered the 1D superlattice Bose Hubbard model at half filling. It has a dimerized Mott insulating phase with topological order characterized by the many-body Zak phase. When the Hamiltonian has inversion symmetry, this many-body Zak phase constitutes a symmetry-protected topological invariant which is strictly quantized. We demonstrated that the many-body Chern number, characterizing a topological Thouless pump in the model \cite{Hayward2018a}, can also be measured using the interferometric scheme. This Thouless pump has recently been realized experimentally \cite{Lohse2015} (see also \cite{Nakajima2016}), but the many-body Chern number was measured only indirectly through transport so far. The scheme discussed here offers a complementary perspective, based on the direct detection of the underlying geometric phases in the many-body wavefunction. 

The method can be generalized to topological superconductors, whose Bogoliubov quasiparticle excitations are characterized by a non-trivial Bloch band topology. For the Kitaev chain, which is in the focus of current research because it hosts isolated Majorana edge states, we demonstrated that the topological Majorana number can be measured directly in the bulk of the system. Such measurements could provide an important step towards a complete understanding of current experiments searching for isolated Majorana edge states \cite{Mourik2012,NadjPerge2014}. 

For the discussion of topological order in spin chains we introduced the spin-Zak phase as a generalization of Hatsugai's local Berry phase \cite{Hatsugai2006} and explained its one-to-one relation to the many-body Zak phase. We showed that topological polarons can be used to detect the topological order of spin chains and gain insights into the pattern of valence bonds in the groundstate. This work can be extended to the investigation of topological order in frustrated magnets and gapped quantum spin liquids using topological polarons.

\section*{Acknowledgements}
The authors would like to thank D. Abanin, M. Aidelsburger, I. Bloch, A. Bohrdt, N. Goldman, M. Hafezi, A. Imamoglu, M. Lohse, F. Pientka, F. Pollmann, C. Schweizer, K. Seetharam, R. Verresen, A. Vishwanath and C. Weitenberg for fruitful discussions. FG thanks 
members of the Graduate School of Excellence MAINZ 
for useful discussions, earlier collaborations and financial support. NYY and EAD thank M. Fleischhauer for useful discussions and earlier collaboration. FG acknowledges financial support from the Gordon and Betty Moore foundation under the EPiQS program, from the Technical University of Munich - Institute for Advanced Study, funded by the German Excellence Initiative and the European Union FP7 under grant agreement 291763, from the DFG grant No. KN 1254/1-1, and DFG TRR80 (Project F8). NYY was supported by the ARO via the Anyon Bridge program under award MURI W911NF-17-1-0323. EAD acknowledges support by Harvard-MIT CUA, AFOSR-MURI: Photonic Quantum Matter (award FA95501610323).

\appendix

\tocless \section{Topological Zak phase of Shiba states}
\label{sec:TopShibaState}

\begin{figure}[t!]
\centering
\epsfig{file=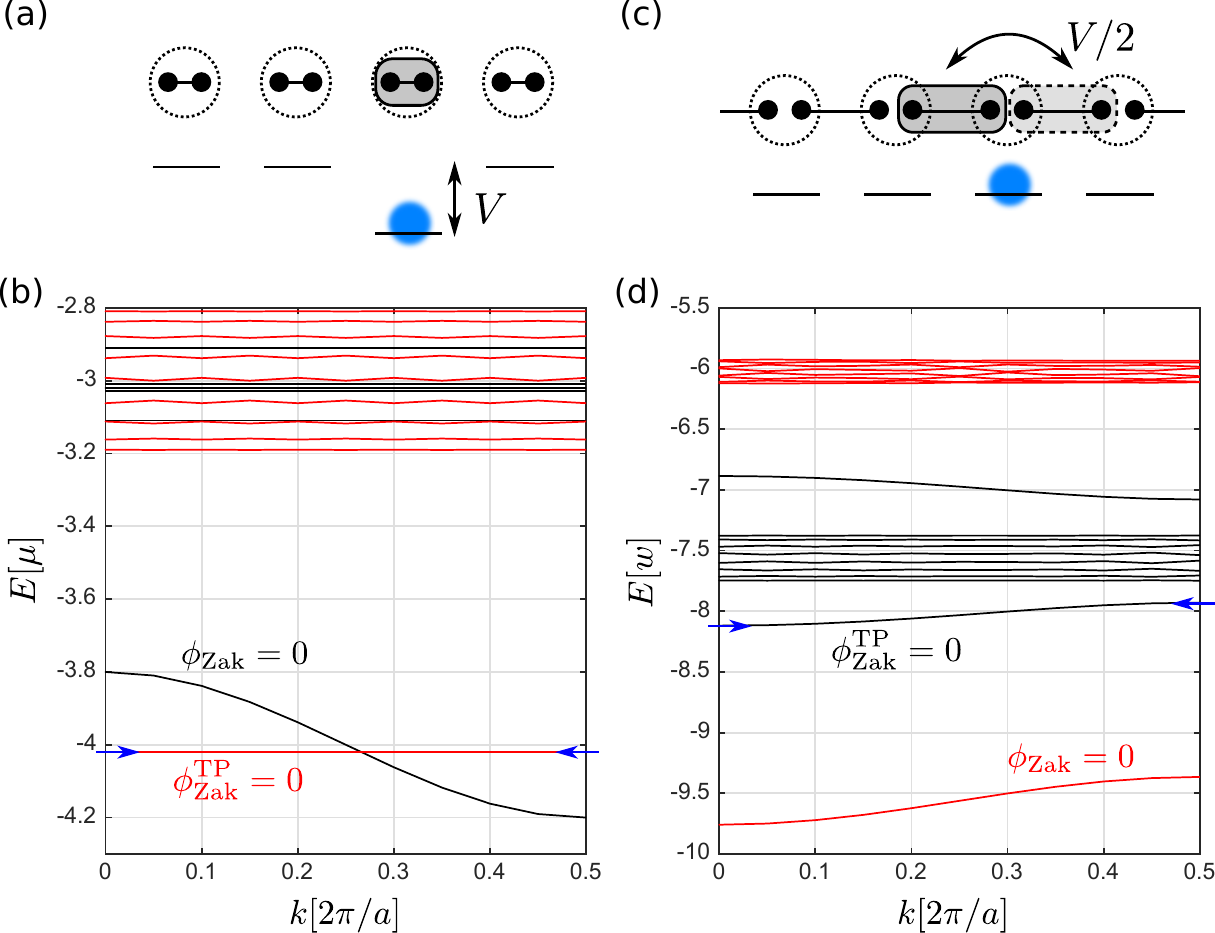, width=0.5\textwidth}
\caption{(a) The TP wavefunction of an impurity (blue) bound to a Bogoliubov quasiparticle (gray) by point-interactions is shown for a topologically trivial case. Black dots correspond to Majoranas, constructed from the original fermions (dashed circles). In (b) the corresponding band structure is shown, assuming impurity-fermion interactions as in Eq.\eqref{eq:HIFlocal}. The fermion parities of the eigenstates are color-coded (red: $P=1$, black: $P=0$), and the TP bound state of interest is marked by blue arrows. Parameters are $w=\Delta=0$, $\mu=1$ for $V=\mu$ and $J=0.1 \mu$. In (c) the TP wavefunction is sketched for a topologically non-trivial case. The corresponding band-structure is calculated in (d) for $w=\Delta$, $\mu=0$ and $V=w$, $J=0.1 w$.}
\label{fig:KitaevTPsV1}
\end{figure}

In this appendix we discuss Shiba states in a $p$-wave superconductor, obtained by assuming local contact interactions between the impurity and the host fermions. Instead of the interaction in Eq.\eqref{eq:HIForiginal} we consider 
\begin{equation}
\H_{\rm IF} =V \sum_{j,\sigma} \cd_{j,\sigma} \c_{j,\sigma} \ad_{j} \a_j.
\label{eq:HIFlocal}
\end{equation}
In the topologically trivial phase, this interaction allows to bind quasiparticle excitations to the impurity, see Fig.\ref{fig:KitaevTPsV1}~(a). The resulting bound state is topologically trivial with the Zak phase $\phi_{\rm Zak}^{\rm TP}=0$.

To understand the topologically non-trivial phase, we consider the limit $\mu = 0$ and $w=\Delta>0$ as in the main text. In terms of the new fermions, see Eq. \eqref{eq:aTildeDef}, the interaction can be expressed as
\begin{multline}
\H_{\rm IF} =\frac{V}{2} \sum_{j,\sigma} \cd_{j,\sigma} \c_{j,\sigma} \biggl[ 1 - \tilde{a}^\dagger_{j} \tilde{a}_{j-1} - \tilde{a}^\dagger_{j-1} \tilde{a}_{j} + \\
+ \tilde{a}_{j-1} \tilde{a}_{j} + \tilde{a}^\dagger_{j} \tilde{a}^\dagger_{j-1} \biggr].
\end{multline}
Let us consider a state with a single Bogoliubov quasiparticle excitation. In the limit where $w \gg V,J$, where $J$ is the nearest-neighbor impurity hopping, we can neglect the pairing terms in the second line. We remain with the first line, which corresponds to an impurity-induced hopping of the Bogoliubov quasiparticle. For $V \gg J$ it is easy to see that this interaction leads to the formation of a topological polaron, where the quasiparticle is bound to the impurity.
 
As sketched in Fig.\ref{fig:KitaevTPsV1} (c), the quasiparticle is delocalized over the two bonds neighboring the impurity in the resulting bound state. The approximate wavefunction in the limit $w \gg V \gg J$ reads
\begin{equation}
\ket{\Phi_{\rm TP}(j)} = \frac{1}{\sqrt{2}} \l \tilde{a}_{j-1}^\dagger + \tilde{a}_{j}^\dagger \r \ket{\psi_{\rm K}} \otimes \cd_j \ket{0},
\end{equation}
when the impurity is localized on site $j$. 

For $J=0$ it is easy to show that this state has a trivial TP Zak phase, $\phi_{\rm Zak}^{\rm TP}=0$, even though the Bogoliubov quasiparticles are topologically non-trivial, $\phi_{\rm Zak}^{\rm K}=\pi$. In contrast to the situation in Fig.\ref{fig:KitaevChainTP} (a), the quasiparticle is delocalized over two consecutive bonds. This non-trivial internal structure of the TP bound state makes it impossible for the impurity to distinguish between quasiparticles residing on the sites (topologically trivial case) and on the bonds (non-trivial case).

For $J>0$ the TP acquires a dispersion, but as long as $J \ll w, V$ the band gap does not close. Therefore the TP Zak does not change and remains trivial, $\phi_{\rm Zak}^{\rm TP}=0$. We confirmed these predictions by exact numerical simulations shown in Fig.\ref{fig:KitaevTPsV1} (b) and (d). Here, as in Fig.\ref{fig:KitaevTPs}, there are two low-lying bands with different fermion parities $P$. One corresponds to an unbound impurity, whereas the second represents a TP bound state. Its Zak phase is always found to be trivial.

\tocless \section{Twisted periodic boundary conditions in spin chains \& many-body spin-Zak phase}
\label{sec:TwistedPBCSpinChain}
In this appendix we show how twisted periodic boundary conditions can be used to classify symmetry-protected topological order in gapped spin systems. We consider models in which the spin along one direction, say $\hat{S}_z$, is conserved, $[\H, \hat{S}_z]=0$. This symmetry is equivalent to particle-number conservation, which we used in Sec.\ref{subsec:MBzakAndTwistedBCs} to construct twisted periodic boundary conditions for band insulators and interacting bosons/fermions.  

\paragraph{Spin $1/2$.--}
First consider a spin-$1/2$ chain, $S=1/2$, with a total number of spins $N$. When $\hat{S}_z$ is conserved we can express the many-body spin wavefunction $\psi_S(x_1,...,x_M)$ describing the spin system in the $\hat{S}_z$ basis. Here $x_j=1,...,N$ denote the coordinates of spin $\uparrow$ particles, where $j=1,...,M$ and $M = N S + S_z^{\rm tot}$ is the number of $\uparrow$-spins. Twisted periodic boundary conditions are defined by imposing
\begin{equation}
\psi_{S}(x_1,...,x_j+N,...,x_M) = e^{i \vartheta} \psi_{S}(x_1,...,x_j,...,x_M)
\label{eq:twistedBCsSpin}
\end{equation}
for all $j=1,...,M$, see Eq.\eqref{eq:twistedBCs}. 

When the twisted periodic boundary conditions are adiabatically changed from $\vartheta=0$ to $2 \pi$, the spin wavefunction $\ket{\psi(\vartheta)}$ picks up a geometric phase up to a gauge transformation $\hat{U}(\vartheta)$. For $\hat{U}(2 \pi)=1$ this gives rise to the following definition of the many-body spin-Zak phase,
\begin{equation}
\phi_{\Zak}^S =  \int_0^{2 \pi} d\vartheta~ \bra{\psi_{S}(\vartheta)} i \partial_\vartheta \ket{\psi_{S}(\vartheta)}.
\label{eq:defPhiSpinZak}
\end{equation}
As in the case of band insulators the value of the spin-Zak phase depends on the gauge choice, see Eqs.\eqref{eq:defPhiZak}, \eqref{eq:defZ}. 

\paragraph{Arbitrary spin.--}
Eq.\eqref{eq:twistedBCsSpin} can be generalized for arbitrary spin $S$ by introducing Schwinger bosons (SB) $\a_j$, $\b_j$ (see e.g. Ref.\cite{Auerbach1998} for an introduction to SBs). When $\hat{S}_z=S_z^{\rm tot}$ is conserved we obtain $\sum_j ( \ad_j \a_j - \bd_j \b_j ) = 2 S_z^{\rm tot}$. Because the total number of SBs is always conserved, $\sum_j ( \ad_j \a_j + \bd_j \b_j ) = 2 N S$, both $\a$- and $\b$-type SB numbers $\sum_j \ad_j \a_j$ and $\sum_j \bd_j \b_j$ are individually conserved. This gives rise to two continuous $U(1)$ gauge symmetries $\mu=a,b$, each of which allows to define a set of twisted periodic boundary conditions. To this end an integer number $n_\mu$ of $\mu$-flux quanta is adiabatically introduced into the system \cite{Laughlin1981} as shown in Fig.\ref{fig:overview} (a). The corresponding unitary operator can be written as
\begin{equation}
\hat{U}(\vartheta_a,\vartheta_b) = \exp \left[ i \sum_{l=1}^N (\vartheta_a \frac{j}{N} \ad_j \a_j + \vartheta_b \frac{j}{N} \bd_j \b_j ) \right].
\end{equation}
Note that we have made a particularly simple gauge choice at this point. As in Eq.\eqref{eq:gaugeChoiceBI} it corresponds to constant forces acting on the SBs respectively. Importantly, for twist angles $\vartheta_\mu = 2 \pi n_\mu$ an integer multiple of $2 \pi$, this corresponds to a pure gauge transformation. We refer to $n_a$, $n_b$ as the integer $S_z$-fluxes.

The most important case for us corresponds to a situation where, say, $\vartheta_b=0$ and $\vartheta_a=\vartheta$. Using $\ad_j \a_j = 2 S - \bd_j \b_j$ we can write
\begin{equation}
\hat{U}(\vartheta) = e^{i \vartheta \sum_{l=1}^N \frac{j}{N} \l \hat{S}_j^z + S \r },
\label{eq:UthetaSpinZak}
\end{equation}
which corresponds to the dynamics generated by a magnetic field gradient across the sample. For $\vartheta = 2 \pi$ this corresponds to a pure gauge transformation and we say that one unit of $S_z$ flux (recall that $n_a$=1) has been introduced into the system. This allows to generalize the definition of the spin-Zak phase to arbitrary $S$:\\

\noindent
\emph{When one unit of $S_z$-flux is adiabatically introduced in a system with periodic boundary conditions (i.e. when the phase $2 \pi S_z$ is picked up by one spin $S_z$ when encircling the system once), the many-body spin wavefunction $\ket{\psi_S}$ returns to itself up to a gauge transformation $\hat{U}$ and a phase, $\ket{\psi_S} \to e^{-i \varphi} \hat{U} \ket{\psi_S}$. The geometric contribution to the phase defines the many-body spin-Zak phase, $\phi_\Zak^S = \varphi - \varphi_{\rm dyn}$.}\\

\noindent
Here the dynamical phase $\varphi_{\rm dyn}$ is defined as the contribution to $\varphi$ which depends on the duration $T$ required for the adiabatic protocol.

For the VB states discussed in the main text, the spin-Zak phase $\phi_\Zak^S$ and Hatsugai's local Berry phase \cite{Hatsugai2006,Hatsugai2007} coincide. Indeed, the phase $\vartheta$ in Eq.\eqref{eq:twistedVBstate} is obtained by applying a magnetic field gradient along the VB. This formalism represents a generalization of Hatsugai's construction where only the complex phase $e^{i \vartheta}$ of the term $J e^{i \vartheta} \hat{S}_i^- \hat{S}_j^+$ was modified "by hand". We have shown here that this procedure can be understood as a direct analogue of the Zak phase for spin systems. 

Hatsugai \cite{Hatsugai2007} discussed a scenario where the gap of the system closes when $\vartheta$ is modified, corresponding to an adiabatic change of the $S_z$ flux in the language used here. Note that the spin-Zak phase is well defined for an arbitrary state with a finite gap $\Delta > 0$ in the thermodynamic limit. In the present case a finite amount of $S_z$ flux can always be eliminated by a gauge transformation in the bulk of the system and hence its bulk gap can not close.

\tocless \section{TPs in spin $S=1$ chains}
\label{sec:HaldaneSpinModel}
\begin{figure}[h]
\centering
\epsfig{file=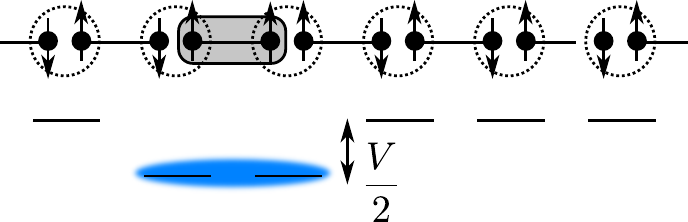, width=0.33\textwidth}
\caption{The TP wavefunction of an impurity (blue) bound to a spin one crackion excitation of the AKLT model is sketched. We expect that this allows to measure the non-trivial spin-Zak phase $\phi_\Zak^S=\pi$ of the AKLT model directly.}
\label{fig:SpinChainTP}
\end{figure}

In this appendix we briefly discuss how TPs can be generalized from the spin-$1/2$ Majumdar-Gosh model to spin-one chains. We start with the AKLT model \cite{Affleck1987}, 
\begin{equation}
\H_{\rm AKLT} = \sum_j P_2(\hat{\vec{S}}_j + \hat{\vec{S}}_{j+1});
\end{equation}
In its groundstate every pair of neighboring spin-one particles has total spin $S=0$ or $S=1$, which is achieved by the projection operator $P_2(\hat{\vec{S}})$ on the total $S=2$ subspace of $\hat{\vec{S}}$. The unique groundstate of the AKLT model is a VB solid which can be understood by writing every spin $1$ as a sum of two spin $1/2$ (and projecting out their singlet sector). This is shown pictorially in Fig.\ref{fig:SpinChainTP}. The presence of VB states gives rise to a non-trivial spin-Zak phase $\phi_\Zak^S=\pi$, as shown by Hirano et al.\cite{Hirano2008} using Hatsugai's method \cite{Hatsugai2006,Hatsugai2007}. Note the close similarity between the AKLT and Majumdar-Gosh models.

To detect the topological order in the AKLT model, we suggest to couple a mobile impurity to an elementary bulk excitation and form a TP. Unlike the groundstate of the AKLT model, its excited states can not be written in closed analytical form. It has been suggested by Knabe \cite{Knabe} that the elementary excitations, termed crackions, correspond to a broken VB dimer and carry spin $S_z=1$, as illustrated in Fig.\ref{fig:SpinChainTP}. Numerical calculations have confirmed that the variational energy of such crackion states is in good agreement with exact results \cite{Garcia-Saez2013}. By coupling an impurity to triplets, in the spirit of Eq.\eqref{eq:defHIS}, we expect that the non-trivial TP spin-Zak phase of crackions should be observable. We expect that a detailed investigation, which we devote to future investigations, can shed new light on the elementary excitations of the AKLT model.

The AKLT model is in the same topological class as the spin-one Haldane model \cite{Haldane1983b}. Therefore we expect that the method can also be applied in this case. TPs in these systems can then be used to explore topological quantum phase transitions, as studied e.g. in Refs. \cite{Hirano2008,Pollmann2012}, experimentally. This includes transitions to topologically distinct phases of the dimerized Heisenberg Hamiltonian \cite{Hirano2008}, or to fully polarized states in the presence of an external magnetic field \cite{Garcia-Saez2013}.


%

\end{document}